\documentclass[twoside,titlepage]{article}
\usepackage{amsmath, amsthm, amssymb} %%%Math enviroments%%%
\usepackage{mathrsfs}
\usepackage{amsfonts}
\usepackage{hyperref} %%%Hyperlinked references enviroment%%%
\usepackage{graphicx} %%%Enviroment for including figures using "includegraphics"%%%
\numberwithin{equation}{section} %%%Sectioning numbering enviroment%%%
\usepackage[margin=10pt,font=small,labelfont=bf,
labelsep=period]{caption}
\usepackage{subfigure}
\usepackage[percent]{overpic}

\newcommand{\be}{\begin{equation}}
\newcommand{\e}{\end{equation}}
\newcommand{\bea}{\begin{equation*}}
\newcommand{\ea}{\end{equation*}}
\newcommand{\la}{\label}

\begin{document}

\begin{titlepage}

%\date{datetext}
\today

\begin{center}

\hfill{\tt WIS/5/11-MAY-DPPA}\\

%\hfill{\tt hep-ph/XXX}\\

\vskip 20mm

{\Large{\bf Near-horizon solutions for D3-branes ending on 5-branes}}

\vskip 10mm

{\bf Ofer~Aharony, Leon~Berdichevsky, Micha~Berkooz and Itamar~Shamir\footnote{Emails: ofer.aharony, leon.berdichevsky, micha.berkooz, itamar.shamir@weizmann.ac.il}
}

\vskip 4mm
{\em Department of Particle Physics and Astrophysics,}\\
{\em Weizmann Institute of Science,}\\
{\em Rehovot 76100, Israel}\\
[2mm]

\end{center}
\vskip 2cm

\begin{center} {\bf ABSTRACT }\end{center}
\begin{quotation}
\noindent
We construct the type IIB supergravity solutions describing
D3-branes ending on 5-branes, in the near-horizon limit of the D3-branes.
Our solutions are holographically dual to the 4d ${\cal N}=4$ $SU(N)$ super-Yang-Mills (SYM)
theory on a half-line, at large $N$ and large 't Hooft coupling, with various boundary conditions that preserve half of the
supersymmetry. The solutions are limiting cases of the general solutions with
the same symmetries constructed in 2007 by D'Hoker, Estes and Gutperle.
The classification of our solutions
matches exactly with the general classification of boundary conditions
for D3-branes ending on 5-branes by Gaiotto and Witten.
We use the gravity duals to compute the
one-point functions of some chiral operators in the ${\cal N}=4$ SYM theory on a half-line at
strong coupling, and find that they do not match with the expectation values of
the same operators with the same boundary conditions at small 't Hooft
coupling. Our solutions may also be interpreted as the gravity
duals of 4d ${\cal N}=4$ SYM on $AdS_4$, with various boundary
conditions.

\end{quotation}

%%%%%%%%%%%%%%
\vfill
%%%%%%%%%%%%%%%%%%
%\flushleft{June 2009}
%%%%%%%%%%%%%%%%%%

\end{titlepage}

\tableofcontents

%\clearpage

\section{Introduction}

Finding the gravitational solutions describing branes intersecting other branes, or branes ending
on other branes, is a challenging problem, which only has a solution in some very special cases.
In addition to its intrinsic interest, this problem is also interesting in the context of the
duality between gravitational theories and quantum field theories, since many quantum field
theories have interesting descriptions using branes ending on other branes (following \cite{Ganor,Hanany:1996ie}), and finding the corresponding gravitational solutions would enable (when they are
weakly coupled and curved) studying the corresponding quantum field theories at strong coupling.

The gravitational solutions for D3-branes intersecting 5-branes along $2+1$ dimensions, in the near-horizon limit of the D3-branes, and in configurations that break half of the supersymmetry of the D3-branes, were found a few years ago in \cite{D'Hoker:2007xy,D'Hoker:2007xz}. In fact, these authors constructed
all solutions that have the same symmetries as this near-horizon limit, whose symmetry algebra is the $2+1$ dimensional ${\cal N}=4$ superconformal algebra $OSp(4|4)$ (with sixteen supercharges). They did this by
finding the general local solution to the BPS equations, and then analyzing an ansatz for the global structure of the solutions. Some of the solutions found in \cite{D'Hoker:2007xy,D'Hoker:2007xz} were conjectured to describe D3-branes intersecting 5-branes, and thus to be dual to the $3+1$ dimensional ${\cal N}=4$ supersymmetric Yang-Mills (SYM) theory with a $2+1$ dimensional defect corresponding to the intersection region (the precise description of this defect depends on the identity of the 5-branes involved). We review the general solutions of \cite{D'Hoker:2007xy,D'Hoker:2007xz} in section \ref{dhetalreview},
and the specific solutions corresponding to D3-branes intersecting 5-branes in section \ref{sec:5brane} (these solutions were also analyzed in detail very recently in \cite{Bachas:2011xa}, which has some overlap with our results in this section).

Configurations of D3-branes ending on 5-branes have the same symmetries as D3-branes intersecting 5-branes, so it is interesting to ask whether they are also included in the solutions classified by \cite{D'Hoker:2007xy,D'Hoker:2007xz}. In section \ref{sec:endingbranes} we answer this question in the affirmative, and show that we can obtain these solutions by a limit of the solutions of D3-branes intersecting 5-branes, in which the number of D3-branes on one side of the 5-branes is taken to zero. From the field theory point of view, the possible boundary conditions for the ${\cal N}=4$ SYM theory that preserve half of the supersymmetry were classified (partly using brane configurations) in \cite{Gaiotto:2008sa,Gaiotto:2008ak}. The solutions we find are dual to the ${\cal N}=4$ SYM theory living on ${\mathbb R}^{2,1}$ times a half-line with such boundary conditions, and we precisely identify the parameters of our solutions with the possible boundary conditions for D3-branes ending on 5-branes, classified in \cite{Gaiotto:2008sa,Gaiotto:2008ak}. Our solutions thus enable for the first time the study of the ${\cal N}=4$ SYM theory living on a half-line with such boundary conditions, in the range of parameters where the gravity solutions are weakly coupled and weakly curved; as usual this range corresponds to taking the large $N$ limit of the ${\cal N}=4$ theory with $SU(N)$ gauge group, with large and fixed 't Hooft coupling $\lambda = g_{YM}^2 N$. This is a necessary first step towards trying to find gravity solutions for more complicated brane configurations, that would involve either D3-branes ending on 5-branes at both ends (for $2+1$ dimensional gauge theories) or D4-branes ending on NS5-branes (for $3+1$ dimensional gauge theories); in both cases the near-horizon limit of the D-branes would not
have a conformal symmetry, so the number of supercharges is at most eight, making it much harder to find the corresponding solutions.

Since ${\mathbb R}^{2,1}$ times a half-line is conformally related to four dimensional anti-de Sitter space ($AdS_4$), our solutions also provide the gravitational dual for the ${\cal N}=4$ SYM theory on $AdS_4$ with the same boundary conditions (at the boundary of $AdS_4$). This problem was recently discussed in \cite{Aharony:2010ay}, where
the gravitational (string theory) duals were found only for boundary conditions related to orbifold or orientifold planes; our analysis completes the discussion of \cite{Aharony:2010ay} by providing the gravitational duals for the theory with more general boundary conditions, coming from D3-branes ending on 5-branes.

Given the solutions for D3-branes ending on 5-branes, we can now perform
computations in the corresponding field theories at large $N$ and large 't Hooft coupling. One thing
which would be nice to compute is the spectrum of these theories (the spectrum of anomalous
dimensions of various operators associated with the defect), but this is a difficult computation,
that was recently discussed (but not completed) in \cite{Bachas:2011xa} for the case of intersecting branes (our solutions are a special case of this). As discussed in \cite{Bachas:2011xa}, this computation is particularly interesting because of the conjecture in \cite{Karch:2001cw} that solutions of this type could exhibit
a ``locally localized graviton'', but we postpone further discussions of this issue to the future. In
section \ref{sec:one-point} we perform the simplest computation in these theories -- that of one-point
functions of chiral operators of ${\cal N}=4$ SYM in the presence of the boundary. On the half-line
with coordinate $z \geq 0$ such one-point functions go like a negative power of $z$ (depending on the
dimension of the operator). We compute the one-point functions of the three lowest-dimension chiral
operators that have non-zero one-point functions, for the general solutions of D3-branes ending
on 5-branes. For the special case of D3-branes ending on D5-branes, which has a weakly coupled limit,
we can compute the one-point
functions also at small 't Hooft coupling, and we find that they differ from the strong coupling
results (the dependence on the specific choice of boundary condition is not the same at weak coupling
as at strong coupling). We end in section \ref{summary} with a summary of our results and a discussion of
remaining open questions.

\section{Review of type IIB solutions with $OSp(4|4)$ symmetry}
\label{dhetalreview}

\subsection{Type IIB supergravity with $ SO(2,3) \times SO(3) \times SO(3) $ symmetry}

In \cite{D'Hoker:2007xy,D'Hoker:2007xz}, all type IIB supergravity
solutions invariant under the 3d ${\cal N}=4$ superconformal group $OSp(4|4)$
(which contains 16 supercharges and the bosonic symmetry $SO(2,3)\times SO(3)\times SO(3)$)
were obtained by D'Hoker, Estes and Gutperle. The main motivation for this
work was to find AdS/CFT duals of 4d $\mathcal{N}=4$ super Yang-Mills theory
with a 3d defect and maximal supersymmetry, but these are also precisely the symmetries
of D3-branes ending on 5-branes, in the near-horizon limit of the D3-branes.
The symmetries above suggest a space-time manifold which is a warped product
\begin{equation}
AdS_{4}\times S_{1}^{2}\times S_{2}^{2}\times\Sigma,\label{eq:space-time manifold}
\end{equation}
where $\Sigma$ is an orientable Riemann manifold over which the $AdS_{4}$
and the two spheres are warped.

We adopt the conventions of D'Hoker et al. (see \cite{D'Hoker:2007xy} and references therein).
We join
the real NS-NS 3-form $H_{(3)} = d B_{(2)}$
and the real RR 3-form $F_{(3)} = d C_{(2)}$ into a complex 3-form field strength ${\tilde F}_{(3)}=d{\tilde B}_{(2)}$ according to
\begin{equation}\label{complex 3-form field strength}
{\tilde F}_{(3)}=H_{(3)}+i F_{(3)},\qquad\qquad {\tilde B}_{(2)} = B_{(2)} + i C_{(2)},
\end{equation}
and the self-dual 5-form field strength is related to the real 4-form potential $C_{(4)}$ by
\begin{equation}\label{fiveformdef}
F_{(5)}  =  dC_{(4)}+\frac{i}{16}({\tilde B}_{(2)}\wedge\bar{\tilde F}_{(3)}-\bar{\tilde B}_{(2)}\wedge {\tilde F}_{(3)}).
\end{equation}

We write a general ansatz for the bosonic fields
that is compatible with the symmetries. For the metric, allowing for
warp factors of the $AdS_{4}$ and $S_{1,2}^{2}$ over the Riemann surface, we have
\begin{equation}
ds^{2}=f_{4}^{2}ds_{AdS_{4}}^{2}+f_{1}^{2}ds_{S_{1}^{2}}^{2}+f_{2}^{2}ds_{S_{2}^{2}}^{2}+ds_{\Sigma}^{2}.
\label{eq:space-time metric ansatz}
\end{equation}
Here $f_{1,2,4}$ are real functions on $\Sigma$.
We can always choose complex coordinates $w$ on $\Sigma$ such that
\begin{equation}
ds_{\Sigma}^{2}=4\rho^{2}\left|dw\right|^{2},
\end{equation}
for some real function $\rho$ on $\Sigma$. As in \cite{D'Hoker:2007xy}, hatted vielbeins $\hat{e}$ will refer
to orthonormal frames in the product space \eqref{eq:space-time manifold},
while unhatted ones will refer to vielbeins of the full space-time geometry
\eqref{eq:space-time metric ansatz}. Thus
\begin{eqnarray}
e^{m} & = & f_{4}\hat{e}^{m},\,\,\,\,\,\,\,\,\,\,\,\,\,\,\,\,\,\,\,\,\,\,\,\, m=0,1,2,3,\nonumber \\
e^{i_{1}} & = & f_{1}\hat{e}^{i_{1}},\,\,\,\,\,\,\,\,\,\,\,\,\,\,\,\,\,\,\,\,\,\,\,\, i_{1}=4,5, \label{vielbeins}\\
e^{i_{2}}  &= & f_{2}\hat{e}^{i_{2}},\,\,\,\,\,\,\,\,\,\,\,\,\,\,\,\,\,\,\,\,\,\,\,\, i_{2}=6,7. \nonumber
\end{eqnarray}
The most general non-trivial 2-form potential that is compatible with the symmetries is
\begin{equation}
{\tilde B}_{(2)}=b_{1}\hat{e}^{45}+ib_{2}\hat{e}^{67},
\label{eq:ansatz for 2-form potential}
\end{equation}
where $b_{1,2}$ are complex functions on $\Sigma$. Similarly, for
the self-dual 5-form
\begin{equation}
F_{(5)}=f_{a}(-e^{0123a}+\varepsilon^{ac}\delta_{cb}e^{4567b}),
\end{equation}
where $a,b=8,9$ are directions along $\Sigma$, and $f_{a}$ are real
functions on $\Sigma$. It is convenient to define
\begin{equation}
e^{z}=\frac{1}{2}(e^{8}+ie^{9})=\rho dw,
\end{equation}
and then using \eqref{vielbeins} we can write
\begin{equation}
F_{(5)}=-2{\rm Re}(f_{z}\rho dw)f_{4}^{4}\hat{e}^{0123}+2{\rm Im}(f_{z}\rho dw)f_{1}^{2}f_{2}^{2}\hat{e}^{4567}.
\end{equation}

\subsection{Local solutions of the BPS equations}

Solutions of the supergravity equations of motion preserving 16 supersymmetries and respecting the bosonic symmetry discussed above, correspond to configurations for which the BPS equations for vanishing of the fermionic fields have 16 independent solutions. Remarkably, D'Hoker et al. were able to crunch the reduced BPS equations into an integrable system and solve it in closed form.
The general solution is given in terms of two real harmonic functions $ h_1 $ and $ h_2 $ on the Riemann surface $ \Sigma $. D'Hoker et al. showed that the $ SL(2,\mathbb{R}) $ symmetry of type IIB supergravity can be used to map any such solution to one where the axion $C_{(0)}$ vanishes, and $b_1$ and $b_2$ are real, so we will assume this from here on for simplicity. That fixes the $SL(2,\mathbb{R})$ symmetry up to the discrete S-duality transformation which reverses the sign of the dilaton and exchanges the two 2-forms; this acts on the solutions by exchanging $h_1$ with $h_2$ and exchanging the two two-spheres.

It is convenient to express the solutions in terms of the following four real functions
\begin{align}
W & \equiv  \partial_{w}h_{1}\partial_{\bar{w}}h_{2}+\partial_{w}h_{2}\partial_{\bar{w}}h_{1}, &
X  & \equiv  i( \partial_{w}h_{1}\partial_{\bar{w}}h_{2}- \partial_{w}h_{2}\partial_{\bar{w}}h_{1}),\\
N_{1} & \equiv  2h_{1}h_{2}\left|\partial_{w}h_{1}\right|^{2}-h_{1}^{2}W, &
N_{2} & \equiv  2h_{1}h_{2}\left|\partial_{w}h_{2}\right|^{2}-h_{2}^{2}W.
\end{align}
The local solutions are as follows \cite{D'Hoker:2007xy,D'Hoker:2007xz}. The dilaton is given by
\begin{equation}
e^{2\Phi} = \frac{N_2}{N_1}.
\label{eq:local solutions - dilaton}
\end{equation}
$W$ obeys $W \leq 0$. The metric factors are
%
%\begin{align}
\begin{equation}
\rho^{2} =  e^{-\frac{1}{2}\Phi}\frac{\sqrt{N_{2}\left|W\right|}}{h_{1}h_{2}},\quad
f_{1}^{2} =  2e^{\frac{1}{2}\Phi}h_{1}^{2}\sqrt{\frac{\left|W\right|}{N_{1}}},\quad
f_{2}^{2} =  2e^{-\frac{1}{2}\Phi}h_{2}^{2}\sqrt{\frac{\left|W\right|}{N_{2}}},\quad
f_{4}^{2} =  2e^{-\frac{1}{2}\Phi}\sqrt{\frac{N_{2}}{\left|W\right|}},
\label{eq:local solutions - metric factors}
\end{equation}
%\end{align}
%
and the 2-form potentials are
\begin{equation} \label{eq:2form potentials}
b_1= +2\tilde{h}_2 + 2h_1^2 h_2  \frac{X}{N_1},\qquad
b_2 = -2 \tilde{h}_1 + 2h_1h_2^2  \frac{X}{N_2},
\end{equation}
where $ \tilde{h}_1 $ and $ \tilde{h}_2 $ are the harmonic duals of $ h_1 $ and $ h_2 $, respectively\footnote{That
is, each pair combines to a holomorphic function $ \mathcal{A}(w) = \tilde{h}_1 + i h_1$ and $ \mathcal{B}(w) = h_2 - i \tilde{h}_2$.}.
Note that
$\tilde h_{1,2}$ (and therefore $b_{2,1}$, which can be thought of as the integrals of the two 2-forms fields over the two 2-cycles on which they are non-vanishing) are defined up to additive real
constants. These constants do not affect the 3-form field strengths, but they affect 5-form computations as
we will discuss in detail below. Such a freedom is to
be expected, since in string theory there are large gauge transformations that shift the integral
of 2-form fields over 2-cycles by an integer, and in supergravity we are not sensitive to this
quantization so we have a freedom to perform any shifts.
Finally, the 5-form field strength $F_{(5)}$, for which we will only need the components along $\hat{e}^{4567}$, is given by
\begin{equation}
2{\rm Im}(f_{z}\rho dw)f_{1}^{2}f_{2}^{2}=2{\rm Im}\left(\left[ 3 i(h_{1}\partial_{w}h_{2}-h_{2}\partial_{w}h_{1}) + \partial_{w}(h_{1}h_{2}\frac{X}{W}) \right]\frac{f_{1}^{2}f_{2}^{2}}{f_{4}^{4}} dw \right).
\end{equation}

\subsection{An ansatz for $ h_{1,2} $ using genus $g$ surfaces} \label{subsec:genus g ansatz}

For the solutions above to be regular solutions of type IIB supergravity, we must impose some additional global restrictions on $ h_{1,2} $ as functions on $ \Sigma $. The conditions for such non-singular solutions, whose boundaries are locally $AdS_5\times S^5$, were investigated in \cite{D'Hoker:2007xz}, and can be solved by constructing $ h_{1,2} $ as functions on a hyper-elliptic Riemann surface of genus $ g $.

The genus {\it g} surface can be taken to be the lower-half-plane
\begin{equation}
\Sigma=\{u \in \mathbb{C}\ |\ {\rm Im}(u) \le 0 \},
\end{equation}
characterized by the algebraic equation
\begin{equation}
s^2(u)=(u-e_1)\prod_{k=1}^g(u-e_{2k})(u-e_{2k+1}), \quad e_{i} \in \mathbb R.
\label{eq:algebraic equation of the genus g serfuce}
\end{equation}
Here the $SL(2,\mathbb R)$ symmetry of the lower half-plane\footnote{That is, the conformal Killing group of the lower half-plane, which is not to be confused with the $SL(2,\mathbb{R})$ symmetry of type IIB supergravity, also mentioned before.} was used to fix one of the branch points of $s$ to $e_{2g+2}=\infty$.
The boundary of the Riemann surface $ \partial \Sigma $, which is the real line, will not be a boundary of the full 10d geometry, as we review below.

The ansatz for the holomorphic differentials $ \partial h_{1,2} $ on the surface above is given by
\begin{equation} \label{eq:holo.diff}
%\begin{split}
\partial h_1 = -i \frac{P(u)Q_1(u)}{s^3(u)}du,\qquad\qquad
\partial h_2 = - \frac{P(u)Q_2(u)}{s^3(u)}du,
%\end{split}
\end{equation}
where $P(u)$ is a real polynomial of degree $2g$ with $g$ complex zeros, and $Q_{1,2}(u)$ are real polynomials of degree $g+1$ with real zeros,
\begin{equation}
\begin{split}
P(u)= & \prod_{a=1}^g(u-u_a)(u-\bar u_a), \quad {\rm Im} (u_a) \le 0,\\
Q_1(u)= & \prod_{b=1}^{g+1} (u-\alpha_b), \quad \alpha_{g+1}<\alpha_{g}<...<\alpha_{2}<\alpha_{1} \in \mathbb R,\\
Q_2(u)= & \prod_{b=1}^{g+1} (u-\beta_b), \quad \beta_{g+1}<\beta_{g}<...<\beta_{2}<\beta_{1} \in \mathbb R.\\
\end{split}
\end{equation}
By construction, the holomorphic differentials $\partial h_1$ and $\partial h_2$ have branch points at $u=e_1,\cdots,e_{2g+1}$ and at $e_{2g+2}=\infty$.
It was shown in \cite{D'Hoker:2007xz} that regularity of the solution requires the following ordering for the real roots and branch points: generically we must have\footnote{Less generic cases, when some points coincide, will be discussed below.}
\begin{equation} \label{eq:parameters}
\alpha_{g+1}<e_{2g+1}<\beta_{g+1}<e_{2g}<...<\alpha_{b}<e_{2b-1}<\beta_b<...<e_2<\alpha_1<e_1<\beta_1.
\end{equation}

In the next subsection we will see that the $\{u_a\}$ may be determined in terms of $\{\alpha_b\}$, $\{\beta_b\}$ and $\{e_i\}$. The generic
genus $g$ solution is thus parameterized by $4g+6$ real parameters: $(2g+1)-2$ moduli of $\Sigma$ (we can parameterize them by the values of the $e_i$, after fixing one of them to infinity and two more to some other fixed values by conformal transformations of the half-plane), $2g+2$ real zeros $\{\alpha_b\}$ and $\{\beta_b\}$ of the holomorphic differentials $\partial h_1$ and $\partial h_2$, one overall scale of the dilaton (above we arbitrarily fixed the string coupling to be one at $u=\infty$, but we can always shift the dilaton by a constant), one for the overall scale of the 10-dimensional metric (which again was arbitrarily fixed by our normalization of the various polynomials above), and three coordinates of the global $SL(2,\mathbb{R})$ group of type IIB supergravity that rotate the solutions we wrote to general solutions with a non-vanishing axion field.

\subsection{Asymptotic $ AdS_5 \times S^5 $ regions}

We may understand the topology and geometry of this general solution in three steps. First, we explain how $ S^2_1 $ and $ S^2_2 $ are fibred over $ \partial \Sigma $, and why the points of $ \partial \Sigma $ (except the branch points $\{e_i\}$) are actually regular interior points of the full 10d geometry. The second step is to understand what the solution looks like near each of the branch points. Finally, we will describe the boundary geometry and its interpretation.

For the first step, note that for generic points on $ \partial \Sigma $ not to be a boundary of the full 10d geometry we must demand that the radius of one (and only one) of the spheres will shrink to zero there. That is, generically $ f_1^2=0 $ or $ f_2^2=0 $ (but not both) on $ \partial \Sigma $. This in turn can be shown to imply that either $ h_1 $ or $ h_2 $ should vanish at every point on $ \partial \Sigma $.

\begin{figure}[h!]
  \centering
   \includegraphics[scale=1]{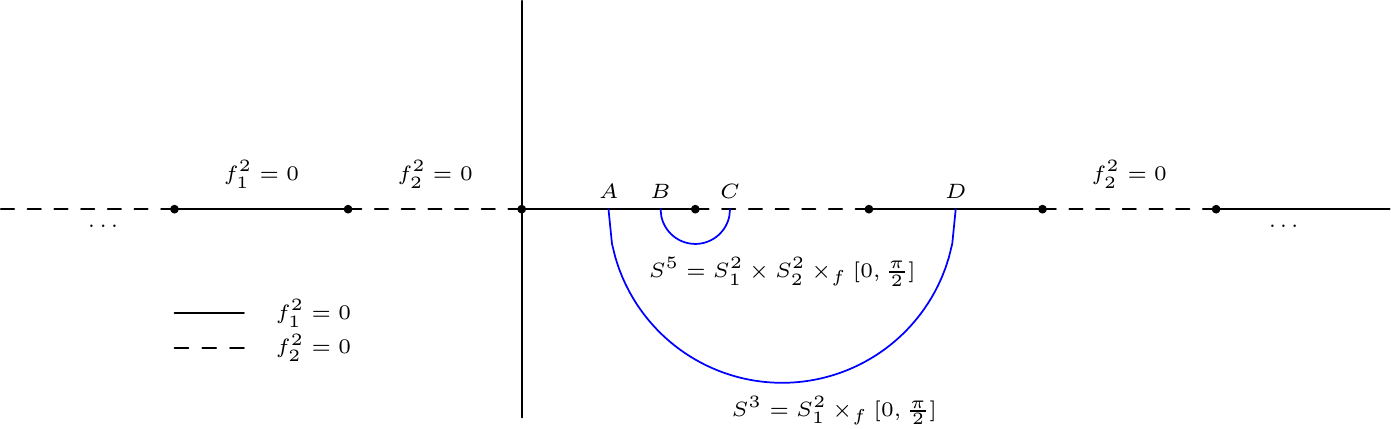}
      \caption{In the hyper-elliptic ansatz, $ \Sigma $ is the lower half of the complex plane. The dots on $ \partial \Sigma $ represent the branch points. On each segment connecting two branch points one (and only one) of the spheres has a vanishing radius. Note that generic points on the real axis $ \partial \Sigma$ are not a boundary of the 10d space-time. }
      \label{fig:u-plane for the general solution}
\end{figure}

It can be seen from \eqref{eq:holo.diff} and \eqref{eq:algebraic equation of the genus g serfuce} that $ \partial h_{1,2} $ alternate at each branch point $e_i$ between taking real and imaginary values on $ \partial \Sigma $. This means that $ h_{1,2} $ satisfy alternating Neumann and Dirichlet boundary conditions along the real line\footnote{This fact can be made transparent by choosing real coordinates $u={\tilde x}+i{\tilde y}$, such that $\partial_u h_j = \frac{1}{2}(\partial_{\tilde x} h_j - i \partial_{\tilde y} h_j).$ }; whenever a branch point $e_i$ on $ \partial \Sigma $ is crossed, the type of boundary condition changes. Moreover, $ h_1 $ and $ h_2 $ have opposite boundary conditions; $ h_1 $ has Neumann boundary conditions whenever $ h_2 $ has Dirichlet, and vice versa (see figure \ref{fig:u-plane for the general solution}). The discussion in the previous paragraph implies in addition that whenever $h_1$ ($h_2$) has a Dirichlet boundary condition, this must take the form $h_1=0$ ($h_2=0$). This is not directly obvious from the ansatz \eqref{eq:holo.diff}; requiring that the values of $h_i$ are the same on each interval where they are supposed to vanish gives $2g$ constraints on the solution, of the form
\be \label{eq:constr}
\begin{split}
\int_{e_{2i}}^{e_{2i-1}} (\partial_u h_1 du +  \partial_{\bar u} h_1 d{\bar u})=0, \quad\quad \int_{e_{2i+1}}^{e_{2i}} (\partial_u h_2 du + \partial_{\bar u} h_2 d{\bar u})=0; \quad i=1,\cdots,g,
\end{split}
\e
which can be used to fix the values of $\{u_a\}$.

The discussion above tells us what cycles we have in the full 10d geometry. Curves whose end points are on $ \partial \Sigma $ and are separated by at least one branch point lead to a non-trivial cycle. For example, in figure \ref{fig:u-plane for the general solution}, consider the curve joining the points $ B $ and $ C $, which we may take to be parameterized by $ y \in [0,\pi/2] $. At the point $ B $ ($ y=\pi/2 $) $ S_1^2 $ has a vanishing radius, while at the point $ C $ ($ y=0 $) $ S_2^2 $  does. Hence, this gives a 5-cycle in the full geometry. We will indeed see below that near each branch point the compact part of the geometry will have the topology of $ S^5 $. By similar arguments, the curve joining $ A $ and $ D $, jumping over two branch points, gives a 3-cycle, since the same sphere ($ S_1^2 $) vanishes on both ends of the curve.

Next we want to describe how the solution looks near the branch points $\{e_i\}$. We will show that the (non-compact)
Riemann surface $\Sigma$ develops a semi-infinite spike there, corresponding to an
asymptotically $AdS_5\times S^5$ region. Choosing coordinates $v=u-e_i$ ($v=-1/u$ for the branch point at $\infty$), the real harmonic functions $h_{1,2}$ near a branch point $v=0$ assume the form
\begin{equation}\label{leadingh12}
\begin{split}
h_1 = & 2 i \left( \gamma_1^i \frac{1}{\sqrt{v}} - \delta_1^i \sqrt{v} \right) + O(v^{3/2}) + c.c.,\\
h_2 = & 2  \left( \gamma_2^i \frac{1}{\sqrt{v}} - \delta_2^i \sqrt{v} \right) + O(v^{3/2}) + c.c.,
\end{split}
\end{equation}
where the constants are
\begin{equation}
%\begin{split}
\label{eq:gamma_1,2 for general genus solution}
\gamma_1^i = \frac{P(e_i)Q_1(e_i)}{\Pi_{j \neq i} (e_i-e_j)^{3/2}},\qquad\qquad
\gamma_2^i = \frac{P(e_i)Q_2(e_i)}{\Pi_{j \neq i} (e_i-e_j)^{3/2}},
%\end{split}
\end{equation}
and
\begin{equation}
\begin{split}\label{eq:delta_1,2 for general genus solution}
\delta_1^i = & \gamma_1^i \left [ \sum_{k=1}^g \left( \frac{1}{e_i-u_k}+\frac{1}{e_i-\bar{u}_k} \right) + \sum_{k=1}^{g+1} \frac{1}{e_i-\alpha_k} - \frac{3}{2} \sum_{k \neq i}^{2g+1} \frac{1}{e_i-e_k} \right],\\
\delta_2^i = & \gamma_2^i \left[ \sum_{k=1}^g \left( \frac{1}{e_i-u_k}+\frac{1}{e_i-\bar{u}_k} \right) + \sum_{k=1}^{g+1} \frac{1}{e_i-\beta_k} - \frac{3}{2} \sum_{k \neq i}^{2g+1} \frac{1}{e_i-e_k}  \right],\\
\end{split}
\end{equation}
for $i=1,\ldots,2g+1$. Note that $\gamma_a^i$ and $\delta_a^i$ ($a=1,2$) alternate
between real and imaginary values as we go along the boundary. At $ u=\infty $ we have $\gamma_1^\infty=\gamma_2^\infty=i$, and
\begin{equation}
\begin{split}\label{eq:delta_1,2 at inf for general genus solution}
-i \delta_1^\infty = & \sum_{k=1}^g \left( u_k+\bar{u}_k \right) + \sum_{k=1}^{g+1} \alpha_k - \frac{3}{2} \sum_{k =1}^{2g+1} e_k,\\
-i \delta_2^\infty = & \sum_{k=1}^g \left( u_k+\bar{u}_k \right) + \sum_{k=1}^{g+1} \beta_k - \frac{3}{2} \sum_{k =1}^{2g+1} e_k.
\end{split}
\end{equation}

In terms of real coordinates $x$ and $y$ defined by $v=e^{-2(x+iy)}$, with $- \infty \le x \le \infty$ and $ 0 \le y \le \pi/2$, the asymptotic region $v \rightarrow 0$ maps to $x \rightarrow \infty$. In this limit, the dilaton behaves as
\begin{equation}
e^{\Phi}= \left|\frac{\gamma_2^i}{\gamma_1^i}\right|+O(e^{-4x}).
\end{equation}
The metric factors for the half of the branch points for which $\gamma_a^i$ and $\delta_a^i$ are real are, to leading order,
\begin{align}\label{metricads}
\rho^2 = & 2 \sqrt{2 |\Delta_i|} + O (e^{-2x}),&
f_1^2 = & 8 \sqrt{2 |\Delta_i|} \sin^2 (y)  + O (e^{-2x}),\nonumber\\
f_2^2 = & 8 \sqrt{2 |\Delta_i|} \cos^2 (y)  + O (e^{-2x}),&
f_4^2 = & 8 \frac{|\gamma_1^i| |\gamma_2^i|}{\sqrt{2 |\Delta_i|}} e^{2x} + O(1),
\end{align}
where  $\Delta_i\equiv \gamma_1^i \delta_2^i -\gamma_2^i \delta_1^i$. Note that here $\rho^2$ is the coefficient in the metric of $4(dx^2+dy^2)$. The other branch points have similar expressions with $f_1^2$ and $f_2^2$ interchanged. The coordinate $ y $ here is precisely the one which traverses the circular curve from $ B $ to $ C $ depicted in figure \ref{fig:u-plane for the general solution}. We can see that, indeed, $ f_{1,2}^2 $ and the $dy^2$ term in the metric of $\Sigma$ combine in the correct way to give a 5-sphere close to the singular points. Likewise, for $ x \rightarrow \infty $, $AdS_4$ with the warp factor $f_4^2$  joins the $x$ coordinate to give an $ AdS_5 $ (up to corrections of order $e^{-2x}$),
\begin{equation}
ds^2_{AdS_5} = dx^2 + \cosh^2(x) ds^2_{AdS_4}.
\end{equation}
The geometry is thus asymptotically $AdS_5 \times S^5$.
The 5-form flux over the 5-sphere near the singular points can be shown to have the expected value for an
$AdS_5\times S^5$ solution with the metric \eqref{metricads} above\footnote{We adopt the conventions of \cite{Grana:2005jc} for the fluxes, with the only difference being a factor of 4 in our definition of the self-dual 5-form field strength \eqref{fiveformdef}. Note that at the classical level the fluxes are defined up to an overall normalization constant in the supergravity action, which is fixed once we fix the normalization of one of the fluxes.},
\begin{equation}
 \lim_{x \rightarrow \infty} \int_{S^5} F_{(5)} =  2 (4 \pi)^3 \Delta_i, \quad \textrm{where} \quad N_i = 8 (4 \pi)^3  |\Delta_i| \in \mathbb{Z}
\label{5-form flux in the asymptotic region}
\end{equation}
is (after charge quantization is taken into account) the integer D3-brane charge. Using
the fact that $W \leq 0$, one can show that $\Delta_i$, and thus the
5-form flux, alternates between positive and negative values (depending on
whether the $\gamma_a^i$ and $\delta_a^i$ are real or imaginary). One can
also show that the total 5-form flux summed over all singular points
vanishes, as expected.

All the information extracted above regarding the $ AdS_5 \times S^5 $ geometry is captured in the two leading orders of $ h_{1,2} $ (or analogously of $ \partial h_{1,2} $) in the expansion around the branch points \eqref{leadingh12}. This will be a recurring theme in this paper. Below, we will see other types of singularities, giving rise to different geometries. We note, from \eqref{eq:holo.diff}, that in this generic case the leading singularity in both $ \partial h_1 $ and $ \partial h_2 $ is $ 3/2 $, i.e. $ v^{-3/2} $. We denote this by $ (3/2,3/2) $, where the first entry refers to the singularity in $ \partial h_1 $ and the second to that of $ \partial h_2 $.
Clearly, the degree of the singularity depends on the coordinate system. It is to be understood that when we use the notation above for the singularities in $ \partial h_{1,2} $ we always refer to the original coordinates in \eqref{eq:holo.diff}.

The general genus $g$ solution thus has $(2g+2)$ asymptotic $AdS_5\times S^5$ regions, with 5-form fluxes proportional to $N_i$.
In the dual field theory this seems to describe $(2g+2)$ half-lines, with some 4d ${\cal N}=4$
$SU(N_i)$ SYM theory living on each half-line, joining together and interacting at a point (times ${\mathbb R}^{2,1}$). To see this, consider the boundary of
our space-time (\ref{eq:space-time metric ansatz}). Writing the $AdS_4$ metric in Poincar\'e coordinates as $ds_{AdS_4}^2 = (dx^{\mu} dx_{\mu} + dz^2) / z^2$ ($\mu=0,1,2$, $z > 0$), there is a boundary wherever the coefficient of $dx^{\mu}dx_{\mu}$ diverges. One place where this happens is where $f_4$ diverges. This happens precisely at the branch points $\{e_i\}$, and we saw
that at each such point we have an $AdS_5\times S^5$ geometry, with a radial coordinate $x$, and with a
four dimensional boundary given by the half-line $\{x^{\mu}, z > 0\}$. (We can view this boundary either
as a half-line, or as an $AdS_4$, the two are related by a conformal transformation.) Another component
of the boundary is at $z=0$. This component is three dimensional, parameterized just by $\{x^{\mu}\}$.
The radial coordinate
approaching this boundary component is $z$, and the other coordinates, whose size remains fixed
as we approach the boundary, include the full Riemann surface $\Sigma$, as well as the two
two-spheres. Thus, all the half-lines discussed above end on this 3d boundary component, which can be
viewed as the intersection of $(2g+2)$ half-lines.

The generic solutions with $g > 0$ that have interior zeros $u_a$ are actually singular at
$u=u_a$ (the metric of $\Sigma$ has a conical singularity there), and we will not discuss them further here\footnote{We thank C. Bachas and J. Estes for bringing this point, which was mentioned in a footnote in \cite{Bachas:2011xa}, to our attention.}. Thus, the only generic solutions that
are non-singular are the genus zero ``Janus solutions''. However, we will see in the next
section that the higher genus solutions have a particular degeneration limit in which they
give sensible solutions of type IIB string theory.

\section{Solutions with 5-branes} \label{sec:5brane}

The generic solutions of the previous section have various degeneration limits, where
several branch points and/or zeros of $Q_i$ come together. One interesting limit, discussed in
\cite{D'Hoker:2007xz}, is when two adjacent branch points $e_i$ come
together; this limit describes 5-branes. Between every two branch points
there is one $\alpha_i$ or one $\beta_i$, so either an $\alpha$ point or a $\beta$ point also
joins the two branch points in this degeneration.

We have already emphasized that the information about the asymptotic $ AdS_5 \times S^5 $ geometry is encoded in the singular behavior of the differentials $\partial{h_{1,2}}$ near the branch points. The same is also true for the 5-branes at the degeneration point. For this reason, we may find the properties of the 5-branes from the simplest possible case of genus one, without any sacrifice of generality.

\subsection{The genus one case} \label{subsec:genus1}

\begin{figure}[h!]
  \centering
   \includegraphics[scale=1.2]{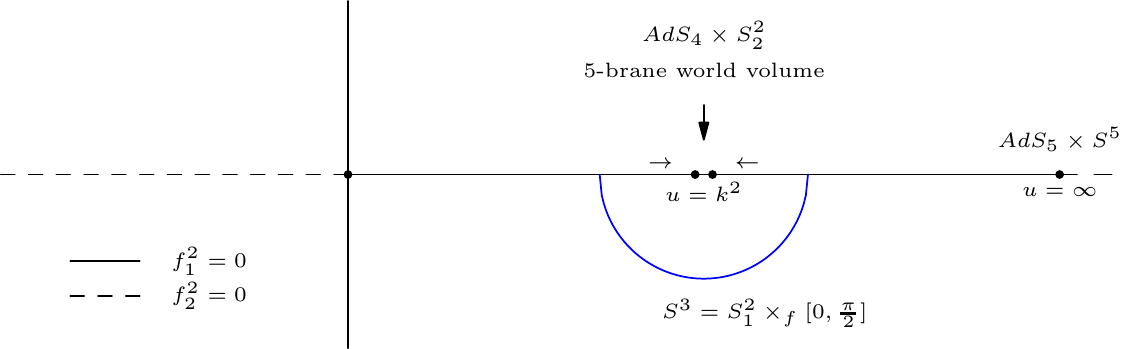}
      \caption{The generic genus one solution has four asymptotic regions. The 5-brane solution is achieved by collapsing two such adjacent regions. The picture shows that in the full geometry there is a 3-cycle surrounding the object at $u=k^2$, as expected of a 5-brane.}
      \label{fig:genus 1 in the collapse limit}
\end{figure}

The generic genus one solution has four branch points at $u=e_{1,2,3},\infty$. There are four additional parameters: $ \alpha_{1,2} $ which are the two real zeros of $ \partial h_1 $, and $ \beta_{1,2} $ which are those of $\partial{h_2}$. They satisfy the ordering\footnote{Here, and throughout this paper,
we arbitrarily choose the branch point at infinity to come between a $\beta$ point and an $\alpha$ point, in this order along the boundary. Solutions with the opposite order may easily be found from the ones we describe by an S-duality transformation, or by taking $u \to -{\bar u}$.}
\begin{equation}
\alpha_2 < e_3 < \beta_2 < e_2 < \alpha_1 < e_1 < \beta_1.
\end{equation}
Following \cite{D'Hoker:2007xz} we choose the collapse $e_1=e_2 \equiv k^2$ (with $k>0$), which implies $\alpha_1=k^2$. We may also fix $e_3=0$, and relabel $\alpha_2=\alpha$ (see figure \ref{fig:genus 1 in the collapse limit}). The four remaining parameters are
\begin{equation}
\alpha<0<\beta_2<k^2<\beta_1.
\end{equation}
Recall that $ \partial h_{1,2} $ also have a mutual complex zero $ u_1 $, which is not free and should be fixed as a function of the other parameters. It is shown in \cite{D'Hoker:2007xz} that regularity and the negativity of $W$ imply that in our limit $ u_1 = k^2 $. We then have
\begin{equation}\label{genusone}
\partial h_1 = -i \frac{(u-\alpha)}{u^{3/2}}du,\qquad\qquad
\partial h_2 = - \frac{(u-\beta_1)(u-\beta_2)}{(u-k^2)u^{3/2}}du.
\end{equation}
As before, there is a $ (3/2,3/2) $ singularity at $ u=0,\infty $, corresponding to asymptotic $ AdS_5 \times S^5 $ regions. However, the degeneration limit has resulted in a new type of singularity  $ (0,1) $ at $ u=k^2 $.

The behavior near the branch points at $ u=0,\infty $ is fully described by $ \gamma^{0,\infty}_{1,2} $ and $ \delta^{0,\infty}_{1,2} $ from \eqref{eq:gamma_1,2 for general genus solution}, \eqref{eq:delta_1,2 for general genus solution}. In terms of the parameters above, they are given by
\begin{align}
\gamma_1^0 & = \alpha, &  \gamma_1^{\infty} & = i, \nonumber \\
\gamma_2^0 & = \frac{\beta_1 \beta_2}{k^2}, &  \gamma_2^{\infty} & = i, \nonumber \\
\delta_1^0 & = -1, &  \delta_1^{\infty} & = i \alpha, \\
\delta_2^0 & = \frac{\beta_1 \beta_2-k^2(\beta_1+\beta_2)}{k^4}, &  \delta_2^{\infty} & = i (\beta_1+\beta_2-k^2), \nonumber
\end{align}
such that
\begin{align}\label{eq:genus 1 - 5-flux at the asymptotic regions}
\Delta_0 & = \frac{k^2 \beta_1 \beta_2 + \alpha \beta_1  \beta_2 -\alpha k^2(\beta_1+\beta_2)}{k^4}, &  \Delta_{\infty} & = \alpha+k^2-\beta_1-\beta_2.
\end{align}
Note that the $SL(2,\mathbb R)$ symmetry of type IIB supergravity was used to set the string coupling near $e_{2g+2}=\infty$ to one. We see that three of the parameters of our solution correspond to the D3-brane charges in the two remaining asymptotic regions, proportional to $\Delta_0$ and $\Delta_{\infty}$, and to the string coupling at the second asymptotic region $e_3=0$.
Choosing a new coordinate $ w^2=u $, such that $ \Sigma = \{ {\rm Re}(w)\le 0, {\rm Im}(w)\ge 0 \} $,
the differentials \eqref{eq:holo.diff} can be integrated in closed form (see \cite{D'Hoker:2007xz})
\begin{equation}
\label{eq:h_1,2 in the collapse limit}
\begin{split}
h_1 = & - 2 i \left( w - \bar{w}  \right) \left[1-\frac{\alpha}{|w|^2} \right],\\
h_2 = & - 2  \left( w + \bar{w} \right) \left[1+\frac{\beta_1 \beta_2}{k^2|w|^2} \right] + \frac{(\beta_1-k^2)(k^2-\beta_2)}{k^3} \ln \left( \frac{|w-k|^2}{|w+k|^2}\right).\\
\end{split}
\end{equation}

We now zoom in on the singular region near $ w=-k $ by choosing the coordinates $ w = r e^{i\psi}-k $ and expanding in small $ r $. This gives at leading order
\begin{equation}
%\begin{split}
h_1 = 4 c \, r \sin (\psi) ,\qquad\qquad
h_2 = 2 b - 2 d \ln (r^2),\\
%\end{split}
\label{eq:h12 5brane}
\end{equation}
where
\begin{equation}
\begin{split}
c = & 1 - \frac{\alpha}{k^2}, \qquad \qquad
2d = \frac{(\beta_1 - k^2)(k^2 - \beta_2)}{k^3}, \\
b = & 2k +2\frac{\beta_1 \beta_2}{k^3}+\frac{(\beta_1 - k^2)(k^2 - \beta_2)\ln (4k^2)}{k^3} .
\end{split}
\end{equation}

For the dilaton we have at leading order
\begin{equation}
e^{2 \Phi} = \frac{d^2}{c^2} \frac{|\ln(r)|}{r^2},
\end{equation}
such that it diverges as $r\to 0$.
The metric functions at leading order are
\begin{align}\label{eq:metric factors in the collapse limit}
\rho^2 = & 2 \sqrt{cd} \, \frac{1}{r^{\frac{3}{2}} |\ln (r) |^{\frac{1}{4}}}, & f_1^2 = & 8 \sqrt{cd} \, \frac{r^{\frac{1}{2}}}{|\ln (r) |^{\frac{1}{4}}}  \sin^2 (\psi),\nonumber \\
& &  \\
f_2^2 = &  8 \sqrt{cd} \, r^{\frac{1}{2}}|\ln (r) |^{\frac{3}{4}}, & f_4^2 = &  8 \sqrt{cd} \, r^{\frac{1}{2}}|\ln (r) |^{\frac{3}{4}}.\nonumber
\end{align}
Note that $f_4^2$
no longer diverges near the branch points, so $w=-k$ does not give an extra boundary component.
We find that, in the vicinity of $ r=0 $, the metric factorizes to $ AdS_4 \times S^2_2 $ and $ S^3 $,
\begin{equation}
ds^2 = f_4^2(r) \left[ ds^2_{AdS_4}+ds^2_{S^2_2} \right]+ 4 r^2 \rho^2(r) \left[ d\psi^2+ \sin^2(\psi) ds^2_{S^2_1}+\frac{1}{r^2}dr^2 \right].
\end{equation}
Together with the behavior of the dilaton, this suggests an interpretation of the supergravity solution as including NS5-branes, whose world-volume consists of the $AdS_4\times S^2$ at $u=k^2$. Thus, this degenerate limit of the generic solution, in which the point $u_1$
joined with two branch points at the boundary, gives a non-singular configuration
in string theory. The transverse 3-cycle indeed carries 3-form flux, as we now show.

Since the 3-sphere is extended in the directions $ 4,5 $ and $ \psi $, we see from \eqref{eq:ansatz for 2-form potential} that there is only a contribution to the flux coming from $ b_1 $. Recalling that $ b_{1,2} $ are real, we learn that this contribution comes from the real part of the 2-form potential, which in light of \eqref{complex 3-form field strength} gives NS-NS flux. This signals the presence of NS5-branes at the point $ r=0 $.
To evaluate the 3-form flux we need the differentials of the harmonic duals $\tilde{h}_{1,2}$ introduced in \eqref{eq:2form potentials}. Using $\partial \tilde{h}_{1,2}=i\partial h_{1,2}$, and expanding around $w=-k$, we have at leading order $ d \tilde{h}_2 = (4d)d \psi$. The NS-NS 3-form flux is therefore
\begin{equation}
%\begin{split}
\label{nsflux}
%\mathcal H =
\lim_{r\rightarrow 0} \int_{\Sigma_3} H_{(3)} = \lim_{r\rightarrow 0} \int_{S^3} db_1 \wedge \hat e^{45}
=  32 \pi^2 d \equiv n_5, \quad n_5 \in \mathbb{Z};
%\end{split}
\end{equation}
while the RR 3-form flux vanishes. Our solution thus includes $n_5$ NS5-branes wrapping $AdS_4\times S^2$.
The 3-form flux is conserved, and the flux sourced by the 5-branes goes off
into the other regions of the geometry.

To find D5-branes, we should take a different collapse limit such that $ f^2_2 $ vanishes on both sides of the singular point, with a $\beta$ point between the two collapsing branch points. Then, $ S^2_2 $ combines with $ \psi $ to form an $ S^3 $, and we get a contribution from the imaginary part of the 2-form potential, which means that the branes source RR 3-form flux and are therefore D5-branes.
This can also be seen from the S-duality transformation mentioned above.

The computation of the 5-form flux near the 5-brane ``throat'' is complicated by the presence of a Chern-Simons type term. The 5-form $F_{(5)}$ \eqref{fiveformdef} is not conserved, but rather satisfies (in our normalizations)
$d F_{(5)} = {1\over 4} H_{(3)} \wedge F_{(3)}$. Thus, if we want a
conserved 5-form that could lead to a conserved charge, we need to take a different 5-form
\be \label{tildefive}
{\tilde F}_{(5)} = F_{(5)} + a C_{(2)} \wedge H_{(3)} - \left({1\over 4} -a \right) B_{(2)} \wedge F_{(3)},
\e
for
some real constant $a$. Generally the Page charge coming from such a 5-form (with $a=0$ or $a=\frac{1}{4}$) is the only conserved and
quantized charge, but it is not gauge-invariant due to the gauge freedom of shifting $B_{(2)}$ and $C_{(2)}$ (see \cite{Marolf:2000cb} for a general discussion). In our solutions we can sometimes fix this freedom by requiring ${\tilde F}_{(5)}$ to be non-singular, even when the two-cycles shrink to zero size.
However, in solutions with NS5-branes like the one we discuss here, the definition \eqref{tildefive} (for generic values of $a$) is not good since there is no globally well-defined non-singular choice of $B_{(2)}$ (and thus of ${\tilde F}_{(5)}$), as we now argue.

The issue is that in solutions with NS5-branes, the $S^2$ on which we have a non-zero $B_{(2)}$ is part of an $S^3$ on which there are $n_5$ units of NS-NS 3-form flux, as we showed above. The $S^2$ shrinks on both poles of the $S^3$, but the value of $B_{(2)}$ (integrated over $S^2$) at these two poles differs by $n_5$. So, if we want $B_{(2)}$ to be globally well-defined, its integral over $S^2$ must be non-zero at least at one
of these poles, but this means that $B_{(2)}$ is singular at that pole. Alternatively, we can define $B_{(2)}$ in different patches (with each patch including a single pole of $S^3$) such that it vanishes at both poles, but then it is not globally well-defined (since we need a non-trivial gauge transformation to relate the different patches; the patches are separated by the generalization of a Dirac string). Similarly, in the presence of D5-branes, there is no globally well-defined and non-singular choice of $C_{(2)}$.

To see this explicitly from our solution, note that to compute $B_{(2)}$ and $C_{(2)}$ we must find the harmonic duals ${\tilde h}_{1,2}$ themselves (rather than just their derivatives). Using again $\partial \tilde{h}_{1,2}=i\partial h_{1,2}$ we have
\begin{equation}\label{eq:harm duals}
\begin{split}
\tilde{h}_1 = &  2 \left( w + \bar{w}  \right) \left[1 + \frac{\alpha}{|w|^2} \right],\\
\tilde{h}_2 = & - 2 i \left( w - \bar{w} \right) \left[1+\frac{\beta_1 \beta_2}{k^2|w|^2} \right] - i \frac{(\beta_1-k^2)(k^2-\beta_2)}{k^3} \ln \left(\frac{w^2 - k^2}{\bar{w}^2 - k^2} \right),\\
\end{split}
\end{equation}
up to arbitrary additive constants.
We see that $b_1=2\tilde{h}_2+\ldots$ changes by $8 \pi d=n_5 / 4\pi$ when going from one side of the point $w=-k$ to the other (on the real line). But the two-cycle $S^2_1$ vanishes on both sides of this point, so we
cannot have $b_1$ vanishing everywhere that the $S^2$ shrinks, and there is no non-singular choice of $B_{(2)}$ (if we want it to be globally well-defined).
%\footnote{when $ \tilde{h}_2 $ is considered a function of the full $ w $ plane}.
This implies that for this solution the only globally well-defined non-singular choice for a conserved 5-form is ${\cal F}_1 \equiv F_{(5)} + {1\over 4} C_{(2)} \wedge H_{(3)}$, where we choose the arbitrary
constant in $\tilde{h}_1$ so that $C_{(2)}$ vanishes everywhere that $S^2_2$ shrinks to zero size (this is a specific fixing of the gauge freedom of shifting $C_{(2)}$). Similarly, in solutions with D5-branes we need to choose ${\cal F}_2 \equiv F_{(5)} - {1\over 4} B_{(2)} \wedge F_{(3)}$.

Thus, in order to compute the 5-form flux in our solution, we need to choose
$\tilde{h}_1$ to vanish on $w \in [0,i\infty)$, as it does for the expression we wrote above. Expanding $\tilde{h}_1$ around $w=-k$ we then find to leading order $\tilde{h}_1=-4k(1+\alpha/k^2)$. The conserved 5-form flux coming from the NS5-brane singularity is therefore
\begin{align}
\begin{split}
\label{eq:5-form flux in the throat}
\lim_{r\rightarrow 0} \int_{\Sigma_3\times S^2_2} (F_{(5)}+{1\over 4} C_{(2)}\wedge H_{(3)}) =& (4 \pi)^2  8 \pi \, 2k(2-c)d \\
=& (4 \pi)^2  8 \pi \frac{(k^2+\alpha)(\beta_1 - k^2)(k^2-\beta_2)}{k^4} \\
=& \frac{1}{4} (N_{\infty} - N_0).
\end{split}
\end{align}
 We see that the 5-form flux going into the 5-brane ``throat'' exactly balances the surplus flux coming from the asymptotic $AdS_5\times S^5$ region at $u=0$ relative to the one at infinity, as it should by charge conservation since $d{\cal F}_1=0$ (note that the second terms in ${\cal F}_1$ and ${\cal F}_2$ do not contribute at the $AdS_5\times S^5$ singularities).

We can now interpret the degeneration limit described above of the genus one solution as describing
the near-horizon limit of D3-branes intersecting NS5-branes, such that some of the D3-branes end on
the 5-branes. More precisely, the four parameters of our solution correspond to the number of
NS5-branes \eqref{nsflux}, the number of D3-branes on each side of it \eqref{eq:genus 1 - 5-flux at the asymptotic regions}, and the relative asymptotic string coupling
between the two asymptotic regions; the near-horizon interpretation is only valid when the string
couplings in the two asymptotic regions coincide.

When our solution has a singularity of this type with $n_5$ NS5-branes or $n_5$ D5-branes wrapping $AdS_4\times S^2$, the low-energy theory on these branes includes a $U(n_5)$ gauge symmetry. The 5-branes intersect the boundary of our space-time along the 3d component of the boundary (where all half-lines intersect), and thus the corresponding field theories have a $U(n_5)$ global symmetry, with the currents localized in three dimensions.

\subsection{D3-branes intersecting several stacks of 5-branes} \label{subsec:3intersect5}

Consider the generic genus $g$ solution reviewed in \S\ref{subsec:genus g ansatz}. The parameters of this solution are the $2g+1$ branch points $e_i$, the $g+1$ real zeros $\alpha_b$ of $\partial h_{1}$, and the $g+1$ real zeros $\beta_b$ of $ \partial h_{2}$.\footnote{Recall that the complex zeros $ u_a $ are not free parameters, and are determined by \eqref{eq:constr}. } It is convenient to represent these parameters as a string of consecutive zeros and branch points by writing (the indices are omitted for simplicity, and the ordering is implicit, see \eqref{eq:parameters})
\begin{equation} \label{eq:para.string}
\alpha \quad e \quad \beta \quad e \quad \alpha \quad e \quad \beta \quad e \quad \ldots \quad e \quad \alpha \quad e \quad \beta \quad e \quad \alpha \quad e \quad \beta.
\end{equation}
As explained above, we may introduce 5-branes into the solution by collapsing adjacent pairs of branch points. When there is an $ \alpha $ between the branch points, we get a stack of NS5-branes. Let us denote such a collapse by a triplet $ (e \alpha e) $. Similarly, D5-branes are obtained by a collapse $ (e \beta e) $.
%
%For simplicity, let us focus on solutions with We are left with $ 2g $ branch points, which should collapse %to a total of $ g $ stacks of 5-branes.
%
Whenever a pair of branch points is collapsed, one of the complex zeros $ u_a $ gets fixed to the collapse point. This results in a $ (0,1) $ singularity for the $ (e \alpha e) $ collapse, and $ (1,0) $ for the $ (e \beta e) $ collapse. As mentioned above, non-singular solutions are
obtained only when there are no points $u_a$ in the interior. Thus, we need to have $ g $ stacks of 5-branes, such that all the $ u_a$ go to the boundary. We then remain with two asymptotic
$AdS_5\times S^5$ regions, as we expect for solutions corresponding to intersecting branes.
We have fixed one of the two $(3/2,3/2)$ branch points to $u=\infty$, and we may fix the other to $u=0$.
\eqref{eq:para.string} has now turned into
\begin{equation} \label{eq:multiple collapse}
\alpha \quad (e \beta e) \quad \alpha \quad \ldots \quad \alpha \quad (e \beta e) \quad \alpha \quad e \quad \beta \quad (e \alpha e)  \quad \beta \quad \ldots \quad \beta \quad (e \alpha e) \quad \beta.
\end{equation}
Clearly, the $ (3/2,3/2) $ branch point at $ u=0 $ has to lie to the left of a $ \beta $ and right of an $ \alpha $ (otherwise there are more than two uncollapsed branch points). There are $ g+1 $ options for how to collapse the other branch points, corresponding to having $ 0,\cdots,g $ $ (e \alpha e) $ collapses (stacks of NS5-branes). Let $ n $ stand for the number of stacks of NS5-branes and $ m $ for the number of stacks of D5-branes. We then have $ n + m = g $. For example, the genus one solution considered above has $ n=1 $ and $ m=0 $.

\begin{figure}[h!]
\subfigure[]{
% {\includegraphics[width=0.45\linewidth,height= 12cm, trim=10 30 35 30, clip]{N3New.eps}}
  {\begin{overpic}[width=0.55\linewidth]{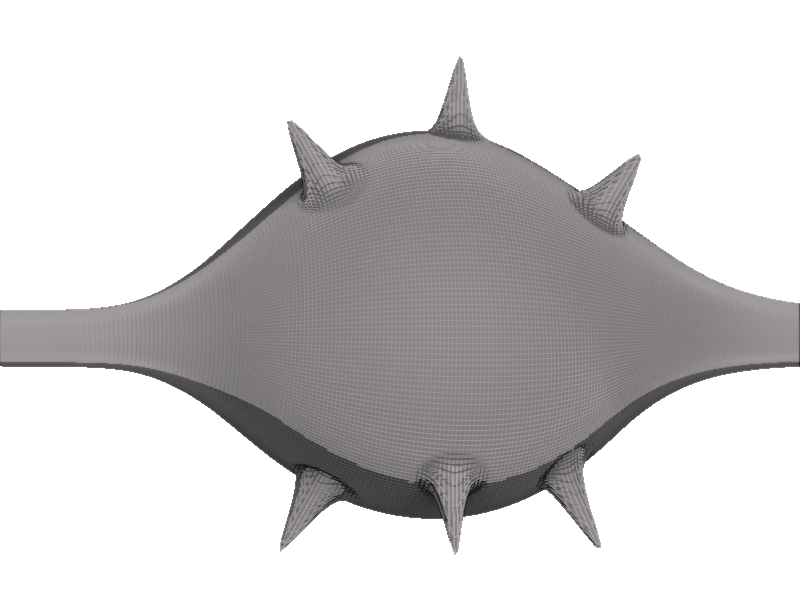}
  \put(-5,40){$AdS_5 \times S_5$}
  \put(90,42){$AdS_5 \times S_5$}
  \put(54,0){D5}
  \put(32,0){D5}
  \put(75,0){D5}
  \put(53,70){NS5}
  \put(32,62){NS5}
  \put(77,58){NS5}
%\put(0,50){$E/\sqrt{T}$}
  \end{overpic}}

}\qquad \qquad
\subfigure[]{
\includegraphics[scale=1.2]{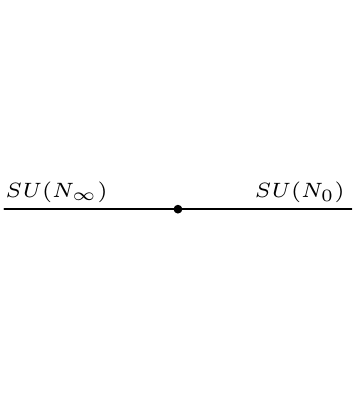}	
}
\caption{(a) A schematic picture of the six dimensional space made from the two
two-spheres and $\Sigma$, for the solutions of this subsection corresponding to D3-branes intersecting D5-branes and NS5-branes. This space is non-compact along
two $AdS_5\times S^5$ ``throats'', and has several D5-brane and NS5-brane ``throats'' coming out of its interior. (b) The dual field theory, which describes two 4d theories on a half-line interacting with a 3d ``defect'' theory. Note that the $z$ coordinate which parameterizes the half-lines in (b) is part of the $AdS_4$ space, and is not visible in (a).}
\label{fig:intersecting branes}
\end{figure}
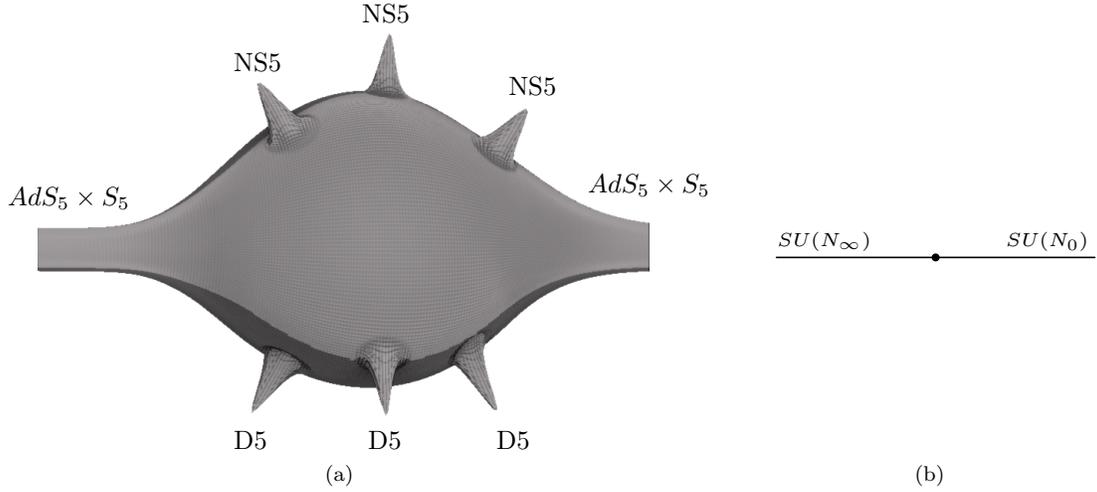

Note that all the NS5-branes are to the right of the $u=0$ branch point, while all the D5-branes are to the left of it. In accordance with this, let us denote the locations of the NS5-branes by $ k^2_a $ ($ a=1,\cdots,n $), and those of the D5-branes by $ -l^2_b $ ($ b=1,\cdots,m $), with $k_a, l_b > 0$. The ordering \eqref{eq:multiple collapse} is then (reinstating the indices)
\begin{equation}
\alpha_{g+1} < -l^2_{m} < \alpha_g < ... < - l_1^2 < \alpha_{n+1} < 0 < \beta_{n+1} < k^2_{n} < ... < \beta_2<k^2_1<\beta_1,
\end{equation}
and the holomorphic differentials are
\begin{equation}
%\begin{split}
\partial h_1  =  - i \frac{\prod_{b=1}^{m+1}(u-\alpha_{b+n})}{\prod_{b=1}^m(u+l_b^2)}\frac{du}{u^{3/2}},\qquad\qquad
\partial h_2 =  - \frac{\prod_{a=1}^{n+1}(u-\beta_a)}{\prod_{a=1}^n(u-k_a^2)}\frac{du}{u^{3/2}}.
%\end{split}
\end{equation}

The 3-form fluxes of each 5-brane stack may be computed as in the previous subsection; we will
discuss the computation of 5-form fluxes in the next section.
We interpret these solutions (in the special case where the asymptotic dilaton is the same in both
$AdS_5\times S^5$ regions) as describing D3-branes intersecting (and possibly ending on) multiple stacks of 5-branes.
The brane orientations are the same as in the standard brane constructions yielding theories with
3d ${\cal N}=4$ supersymmetry \cite{Ganor,Hanany:1996ie}; the NS5-branes fill three of the directions orthogonal
to the D3-branes, and the D5-branes fill the other three orthogonal directions. In the conformal
limit that our solutions describe, all these branes intersect at a point
times $\mathbb{R}^{2,1}$ (before any back-reaction is taken into account). We will discuss the
distinction between the different NS5-brane (D5-brane) stacks in the next section.

\section{D3-branes ending on 5-branes}\label{sec:endingbranes}

Our main goal in this paper is to construct the  supergravity solutions of D3-branes ending on 5-branes. Several years ago, Gaiotto and Witten \cite{Gaiotto:2008sa,Gaiotto:2008ak} classified all possible supersymmetry-preserving boundary conditions for the ${\cal N}=4$ SYM theory, and in particular the boundary conditions that correspond to D3-branes ending on 5-branes, and we would like to compare the results we find to their analysis. In the previous section we looked at solutions that had two asymptotic $ AdS_5 \times S^5 $ regions. In the brane picture, each one of those regions is interpreted as the near horizon limit of a stack of D3-branes, where the number of branes is controlled by the 5-form flux in that region. When the numbers of D3-branes in the two regions are not equal, some of the D3-branes end on 5-branes. Setting one of the 5-form fluxes to zero means that there are no D3-branes in this region, and that the metric there should be that of flat space (instead of having an
$AdS_5\times S^5$ throat). In such a case all the D3-branes coming from the other asymptotic region end
on 5-branes.

We now examine how the gravity solution behaves in this limit. Once again, we first consider the genus one case in full detail, before proceeding to the more general solutions.

\subsection{The genus one case}

Consider the 5-form fluxes $ \Delta _{0,\infty} $ computed in \eqref{eq:genus 1 - 5-flux at the asymptotic regions}. We want to find a limit such that the flux at the origin vanishes
while the flux at infinity is kept finite. Recalling figure \ref{fig:genus 1 in the collapse limit}, the topology suggests that we
must keep the NS5-branes, which are localized at the point $k^{2}$, at a finite distance from the origin. Therefore we shall keep $k$ fixed in this limit. Then, we need to take
\begin{equation}
\alpha,\beta_{2}\rightarrow0.
\label{eq:genus 1 - limit of vanishing flux}
\end{equation}
With this choice
\begin{equation}
\Delta_{0}= 0,\qquad \qquad \Delta_{\infty}= k^{2}-\beta_{1}.
\label{eq:genus 1 - vanishing flux}
\end{equation}
A priori, the correct way to take the limit could involve keeping the ratio of $\alpha$ and $\beta_{2}$ fixed to some value, but we will see that the limit is independent of this ratio.
In the spirit of section \ref{sec:5brane} we think of this limit as a new kind of collapse $ (\alpha e \beta) $.

In this limit, the holomorphic differentials \eqref{genusone} assume the following form
\begin{equation}
\partial h_1 = -i \frac{du}{u^{1/2}},\qquad\qquad
\partial h_2 = - \frac{(u-\beta_1)}{(u-k^2)}\frac{du}{u^{1/2}}.
\end{equation}
We see that this limit gives a new type of singularity $ (1/2,1/2) $ at $u=0$. One can prove
that this is the only other possible singularity of $\partial h_1$ and $\partial h_2$ that can occur as
a limit of the solutions we discuss, without giving a singularity of the full geometry.

Using the exact form of $ h_{1,2} $ given in \eqref{eq:h_1,2 in the collapse limit}, with coordinates $ w = re^{i \theta} $ ($\theta \in[\frac{\pi}{2},\pi]$), near $r=0$ ($u=0$) the real harmonic functions behave as
%
%\begin{eqnarray}
\begin{equation}
h_{1} = 4r\sin (\theta),\qquad\qquad
h_{2} = -\frac{4\beta_{1}r\cos (\theta)}{k^{2}}+\frac{4(k^{2}-\beta_{1})r^{3}\cos(3 \theta)}{3k^{4}}+O(r^{5}).
\label{eq:h after limit}
\end{equation}
%\end{eqnarray}
%
Note that the singular terms drop out.

We can now plug this into \eqref{eq:local solutions - dilaton} and \eqref{eq:local solutions - metric factors} and find the leading behavior near $r=0$
\begin{align}
e^{2\Phi} & = \frac{\beta_{1}^{2}}{k^{4}},&
\rho^{2} & = \frac{2\sqrt{2|\Delta_{\infty}|}}{k^{2}}, \nonumber \\
f_{1}^{2} & =  \frac{8\sqrt{2|\Delta_{\infty}|}}{k^{2}}r^{2}\sin^2 (\theta),&
f_{2}^{2} & = \frac{8\sqrt{2|\Delta_{\infty}|}}{k^{2}}r^{2}\cos^2(\theta),\quad \quad \quad
f_{4}^{2} = \frac{8\beta_{1}}{\sqrt{2|\Delta_{\infty}|}}.
\label{eq:flat metric near the origin after the limit}
\end{align}
Note the $r^{2}$ factor in $f_{1}^{2}$ and $f_{2}^{2}$. This means
that the radius of $S^{5}$ decreases as we approach the branch
point at $w=0$, resulting in a flat metric, as in spherical
coordinates in ${\mathbb R}^6$. Subsequently, there is no topologically non-trivial
cycle, needed to support a 5-form flux. Additionally, the $r^{-2}=e^{2x}$
singularity in $f_{4}^{2}$, responsible for the asymptotic
$AdS_{5}$ structure, is no longer present.

The metric near $w=0$ is therefore
\begin{equation}
ds^{2}=\frac{8\sqrt{2|\Delta_{\infty}|}}{k^{2}}\left(dr^{2}+r^{2} (d\theta^{2}+\sin^2 (\theta) \, ds_{S_{1}^{2}}^{2} + \cos^2(\theta) \, ds_{S_{2}^{2}}^{2})\right)+\frac{8\beta_{1}}{\sqrt{2|\Delta_{\infty}|}}ds_{AdS_{4}}^{2},
\end{equation}
which is precisely the metric of $AdS_{4} \times \mathbb{R}^{6}$.
The point $w=0$ ($u=0$) is thus just a regular point in the full geometry. We have therefore
obtained a smooth solution describing D3-branes ending on NS5-branes, with a single asymptotic $AdS_5\times S^5$
region at $u=\infty$, a single NS5-brane stack located at $u=k^2$, and no
other singular points. Similarly, one may
obtain from a different degeneration limit of the genus one case the solution for D3-branes
ending on D5-branes.

\subsection{D3-branes ending on multiple stacks of 5-branes}

The main lesson of the genus one case is that it is possible to locally turn off the 5-form flux, emanating from an asymptotic $ AdS_5 \times S^5 $ region at $ u=e $, by letting $ \alpha $ and $ \beta $ coalesce to $ e $ (an $ (\alpha e \beta) $ collapse). This changes the singularity at $ e $, from $ (3/2,3/2) $ to $ (1/2,1/2) $, leading to a smooth $ AdS_4 \times \mathbb{R}^6 $ geometry at that point.
A $(\beta e \alpha)$ collapse gives the same result.

The local nature of this procedure means that it is easily applicable to the more general solution of multiple stacks of 5-branes intersecting D3-branes, introduced in \S\ref{subsec:3intersect5}. Consider the schematic representation of this solution given in \eqref{eq:multiple collapse}. Recall that in this solution there are two asymptotic $ AdS_5 \times S^5 $ regions, at $ u=0 $ and $ u=\infty $, corresponding to D3-branes ending on stacks of 5-branes from both sides. Taking an $ (\alpha e \beta) $ collapse at $ u=0 $ leads to a new solution, with D3-branes ending on the 5-branes from only one side :
\begin{equation}
\alpha \quad (e \beta e) \quad \alpha \quad \ldots \quad \alpha \quad (e \beta e) \quad (\alpha e \beta) \quad (e \alpha e)  \quad \beta \quad \ldots \quad \beta \quad (e \alpha e) \quad \beta.
\end{equation}

The remaining $2g$ parameters are
\begin{equation}
\alpha_{g+1}<-l_m^2<\alpha_g<...<\alpha_{n+2}<-l_{1}^2<0<k^2_{n}<\beta_n<...<\beta_2<k^2_1<\beta_1,
\end{equation}
with holomorphic differentials
\begin{equation} \label{eq:nholodiff}
%\begin{split}
\partial h_1  = - i \frac{1}{\sqrt{u}} \prod_{b=1}^{m} \frac{(u-\alpha_{b+n+1})}{(u+l_b^2)}du ,\qquad\qquad
\partial h_2 = - \frac{1}{\sqrt{u}} \prod_{a=1}^{n} \frac{(u-\beta_a)}{(u-k_a^2)}du .
%\end{split}
\end{equation}
It is convenient to substitute $u=w^2$ in $ \partial h_{1,2} $ as in \S\ref{subsec:genus1}. In this coordinate $h_{1,2}$ are given by
\begin{equation}\label{eq:general harm}
\begin{split}
h_1 = &  4 \mathrm{Im}(w) + 2 \sum^{m}_{b=1} \tilde{d}_b \ln \left( \frac{|l_b - i w|^2}{|l_b + i w|^2} \right),\\
h_2 = & -4 \mathrm{Re}(w) - 2 \sum^{n}_{a=1} d_a \ln \left( \frac{|k_a + w|^2}{|k_a - w|^2} \right),\\
\end{split}
\end{equation}
where
\begin{equation}
\begin{split}
d_a \equiv & \frac{(\beta_a-k^2_a)}{2 k_a} \prod_{c \ne a}^{n}\frac{(k^2_a - \beta_c)}{(k^2_a - k^2_c)}, \qquad \qquad
\tilde d_b \equiv \frac{(-\alpha_{b+n+1}-l^2_b)}{2l_b} \prod_{c \ne b}^{m}\frac{(l^2_b + \alpha_{c+n+1})}{(l^2_b - l^2_c)}. \\
\end{split}
\end{equation}
Note that $d_a,\tilde{d}_a>0$ (recall $k_a,l_b>0$).
The coordinate $w$ occupies the second quadrant of the complex plane,
%that is the domain
$\{\mathrm{Re}(w)<0,\mathrm{Im}(w)>0\}$. The NS5-branes are then located on the negative real line ($\{-k_a\}$), while the D5-branes are located on the positive imaginary line ($\{il_b\}$). Near each of these points, the local supergravity solution is as we have discussed in \S\ref{subsec:genus1}. To demonstrate this, expand \eqref{eq:general harm} near a stack of NS5-branes at $w=-k_a$. Using $w=re^{i\psi}-k_a$ we find
\begin{equation}
\begin{split}
h_1  = 4 c_a r \sin(\psi), \qquad \qquad  h_2 = 2 b_a - 2 d_a \ln (r^2),\\
\end{split}
\end{equation}
where $c_a,b_a$ are real constants that depend on the parameters of the solution. This is the same local form of $h_{1,2}$ as we found in \eqref{eq:h12 5brane}. The resulting calculation of the metric and 3-form flux can be carried over without any change. Similar considerations apply to the D5-branes at $w=il_b$.

Repeating the calculations of \S\ref{subsec:genus1},  we find that the 3-form flux carried by the stack of NS5-branes at $w=-k_a$ is given by
\begin{equation}
%\mathcal H_a=
\int_{S^3} H_{(3)} =   n_a, \qquad n_a\equiv 32 \pi^2 d_a \in \mathbb{Z}, \qquad \qquad \int_{S^3} F_{(3)}=0.
\end{equation}
%$\mathcal F_a$.
The number of NS5-branes filling the $AdS_4\times S^2$ at $w=-k_a$ is therefore $n_a$.

A similar analysis near the $b$'th stack of D5-branes ($w = i l_b$) gives
\begin{equation}
\begin{split}
%\mathcal H_b
\int_{S^3} H_{(3)} = 0, \qquad\qquad
%\mathcal F_b
\int_{S^3} F_{(3)} = -m_b, \qquad m_b \equiv 32 \pi^2 \tilde d_b \in \mathbb{Z},
%\quad \Phi_t^b= 4(4\pi)^3l_b m_b,
\end{split}
\end{equation}
such that there are $m_b$ D5-branes in this stack.

\begin{figure}[h!]
\subfigure[]{
% {\includegraphics[width=0.45\linewidth,height= 12cm, trim=10 30 35 30, clip]{N3New.eps}}
   {\begin{overpic}[width=0.55\linewidth]{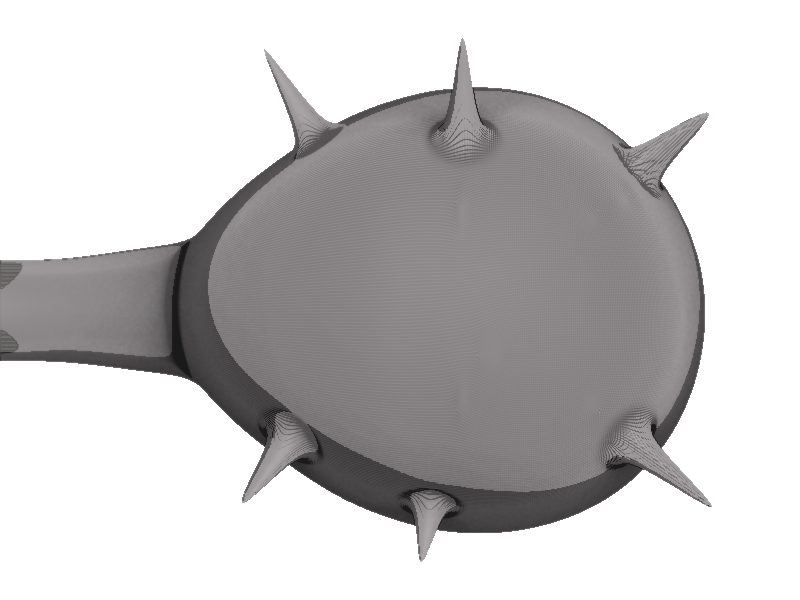}
  \put(50,-1){D5}
  \put(27,6){D5}
  \put(88,6){D5}
  \put(53,72){NS5}
  \put(29,71){NS5}
  \put(85,63){NS5}
\put(-5,47){$AdS_5 \times S_5$}
  \end{overpic}}
}\qquad \qquad
\subfigure[]{
\includegraphics[scale=1.2]{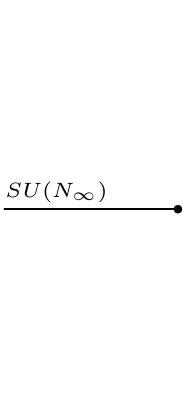}	
}
\caption{(a) A schematic picture of the six dimensional space made from the two
two-spheres and $\Sigma$, for the solutions of this subsection corresponding to D3-branes ending on D5-branes and NS5-branes. This space is non-compact along
 the $AdS_5\times S^5$ ``throat'', and has several D5-brane and NS5-brane ``throats'' coming out of its interior. (b) The dual field theory, which describes the 4d ${\cal N}=4$ SYM theory on a half-line with some boundary condition (that could include interactions with a 3d SCFT at the boundary). As in the previous figure, the 4d boundary component lives infinitely far away along the ``throat'' in figure (a).
}
\label{fig:ending branes}
\end{figure}

We have already discussed the difficulty in defining a conserved and globally well-defined 5-form flux in solutions where there are both NS5 and D5-branes. Before we show how this difficulty may be circumvented, let us first describe the simpler case where only 5-branes of one type appear. As explained towards the end of \S\ref{subsec:genus1}, for solutions that involve only NS5-branes we may use
\begin{equation}
{\cal F}_1 \equiv F_{(5)} + {1\over 4} C_{(2)} \wedge H_{(3)},
\end{equation}
which is both conserved and globally well-defined. Likewise, for D5-branes we use
\begin{equation}
{\cal F}_2 \equiv F_{(5)} - {1\over 4} B_{(2)} \wedge F_{(3)}.
\end{equation}

Hence, for the solutions with only NS5-branes, we find that the 5-form flux penetrating the 5-cycle $S^3 \times S^2_2$ at $w=-k_a$ is given by
\begin{equation}
\int_{\Sigma_5^a} {\cal F}_1 = \frac{1}{4} K_a n_a, \quad K_a \equiv 32 \pi k_a,
\end{equation}
which we interpret as having $n_a K_a$ D3-branes ending on this 5-brane stack, or $K_a$ D3-branes ending on each NS5-brane. Similarly, for the solutions with only D5-branes we find
\begin{equation}
\int_{\Sigma_5^b} {\cal F}_2 = \frac{1}{4} L_b m_b, \quad L_b \equiv 32 \pi l_b,
\end{equation}
with $L_b$ D3-branes ending on each D5-brane. In both cases the total 5-form flux summing over all 5-brane singularities equals the 5-form flux at the $AdS_5\times S^5$ singularity, as expected (all D3-branes end on 5-branes).

 These solutions match nicely with the classification \cite{Gaiotto:2008sa,Gaiotto:2008ak} of the possible half-supersymmetric boundary conditions related to D3-branes ending on D5-branes or NS5-branes. In \cite{Gaiotto:2008sa,Gaiotto:2008ak} the possible boundary conditions for D3-branes ending on D5-branes were discussed by a weak coupling analysis; the direct classification of boundary conditions for D3-branes ending on NS5-branes is more complicated since these involve (in all cases except the case of a single NS5-brane) a non-trivial 3d superconformal field theory (SCFT) on the boundary of the half-line, but it must be the same as that of D5-branes by S-duality. For D3-branes ending on D5-branes,
 the boundary conditions are classified (see \cite{Gaiotto:2008sa,Gaiotto:2008ak} and references therein) in terms of the behavior of three of the adjoint scalar fields $X_i$ ($i=1,2,3$) of the ${\cal N}=4$ SYM theory (the ones corresponding to the motion of the D3-branes along the D5-branes) near the boundary of the half-line at $z=0$. The different boundary conditions correspond to choosing an $N$-dimensional representation $\tau_i$ ($i=1,2,3$) of $SU(2)$ ($[\tau_i, \tau_j] = i \epsilon_{ijk} \tau_k$), and the scalar fields then behave near the boundary as $X_i = \tau_i / z$. Each $N$-dimensional representation can be decomposed into irreducible representations, so that it contains $m_b$ copies of the $L_b$-dimensional representation, and the number of irreducible representations that appears, $\sum_b m_b$, is identified with the number of D5-branes. We interpret such boundary conditions as having $L_b$ D3-branes ending on each of the $m_b$ D5-branes, for every value of $b$, and we thus have the same labeling for our solutions above as for the possible boundary conditions. It is easy to show that the global symmetries $\prod_b U(m_b)$ also agree.

Let us recall the difficulty in finding a conserved and globally well-defined 5-form when there are both D5-branes and NS5-branes. The technical issue is that to define a conserved 5-form we need to have either $B_{(2)}$ or $C_{(2)}$ non-singular. However,
whenever we have a D5-brane singularity, ${\tilde h}_1$ (and, thus, also $C_{(2)}$) jumps by the number of D5-branes as we go along the real line from one side of the D5-brane singularity to the other, so it cannot be taken to vanish all along the region where the corresponding 2-cycle vanishes. The same is true for ${\tilde h}_2$ at NS5-brane singularities. The fact that the definition of the D3-brane charge in this case
is problematic is related to the fact that \cite{Hanany:1996ie} configurations of D5-branes intersecting NS5-branes carry D3-brane charge (due to the Chern-Simons term in the type IIB supergravity action); and,
related to this, the number of D3-branes ending on an NS5-brane (D5-brane) changes as this brane is moved past a D5-brane (NS5-brane), so it is not clear how to identify this number.

\begin{figure}[h!]
  \centering
   \includegraphics[scale=1]{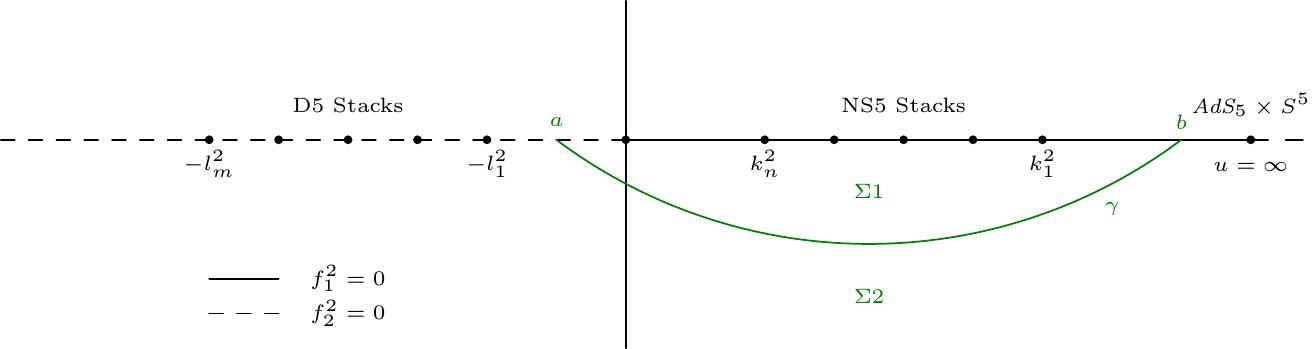}
      \caption{The $u$-plane for solutions of D3-branes ending on NS5-branes and D5-branes, with the $AdS_5\times S^5$ singularity chosen to be at $u=\infty$ and the $AdS_4\times {\mathbb R}^6$ point at $u=0$. We depict the first choice for the surface $\Sigma_1$ on which ${\cal F}_1$ is well-defined, and for the surface $\Sigma_2$ on which ${\cal F}_2$ is well-defined, separated by the curve $\gamma$.}
      \label{fig:flux-definition}
\end{figure}

However, there is a natural way to define a conserved 5-form charge in our solutions for this case as well\footnote{We thank
Don Marolf for this suggestion.}. The 5-form ${\cal F}_1$ is well-defined near all the NS5-brane
singularities at $u=k_a^2$, and the 5-form ${\cal F}_2$ is well-defined near all the D5-brane
singularities at $u=-l_a^2$. We can extend the regions where these two 5-forms are well-defined so
that together they cover all of $\Sigma$. The main constraint is that the region $\Sigma_1$ where
${\cal F}_1$ is well-defined cannot include more than one interval separating different branch points
with negative $u$ (where the $S^2$ on which $C_{(2)} \neq 0$ shrinks to zero size), while the region $\Sigma_2$ where ${\cal F}_2$ is well-defined cannot include more than one
interval separating different branch points with positive $u$ (where the $S^2$ on which $B_{(2)} \neq 0$ shrinks to zero size). This leaves us with two possible choices for
the topology of these regions. We can take $\Sigma_1$ to be a region that intersects the real line
along $[a,b]$, where $-l_{1}^2 < a < 0$ and $k_1^2 < b < \infty$, and $\Sigma_2$ to be the complement
of this region, see figure \ref{fig:flux-definition}; this fulfills the requirements above. There is then a unique non-singular choice for ${\cal F}_1$ in $\Sigma_1$ by choosing $C_{(2)}$ to vanish on $[-l_{1}^2,0]$, and similarly there is a unique
non-singular choice for ${\cal F}_2$ in $\Sigma_2$ by choosing $B_{(2)}$ to vanish on $[k_1^2,\infty]$. The other choice is to take $\Sigma_2$ to be
a region that intersects the real line along $[{\tilde a}, {\tilde b}]$ with $-\infty < {\tilde a} < -l_m^2$ and $0 < {\tilde b} < k_n^2$. The two choices are related by S-duality together with a reflection of
the $u$-plane, so we will focus on the first choice here.

At first sight, the fact that we used two different 5-forms to cover $\Sigma$ does not allow us to
obtain a conserved charge. However, consider the integral of ${\cal F}_1$ on the boundary $\partial \Sigma_1$ of $\Sigma_1$ (times the two two-spheres). Since $d{\cal F}_1=0$, this integral vanishes. On the other hand, it has two contributions; one from the ``external'' boundary of $\partial \Sigma_1$ which is along $\partial \Sigma$, where it gets contributions from the 5-brane singularities analogous to the ones we computed before
(and not from any other points on the boundary), and one from the ``internal'' boundary, along the curve $\gamma$ in figure \ref{fig:flux-definition}. Similarly, the integral of ${\cal F}_2$ on $\partial \Sigma_2$ also vanishes, and it is given by the sum of the
contributions from the singularities along the real line (both the 5-brane singularities and the
$AdS_5\times S^5$ point at $u=\infty$), plus the contribution from the ``internal'' boundary. If we add
these two integrals, the
total contribution from the ``internal boundary'' $\gamma$ is the integral of
\be
{\cal F}_1-{\cal F}_2 = \frac{1}{4} d(B_{(2)}\wedge C_{(2)})
\e
along this boundary; but this is just proportional to the difference in the
values of $B_{(2)}\wedge C_{(2)}$ between the two edges of this boundary $\gamma$ at $u=a$ and $u=b$, and this
vanishes since either $B_{(2)}$ or $C_{(2)}$ vanishes at each of these points. Thus, we find that the
sum of the 5-form fluxes ${\cal F}_1$ or ${\cal F}_2$ over the 5-brane and $AdS_5\times S^5$ singularities vanishes, so this
defines a conserved charge\footnote{A similar conserved charge may be defined for the solutions reviewed in \S\ref{subsec:3intersect5}, describing D3-branes intersecting D5-branes and NS5-branes.}.

Let us now apply this condition to fix $\tilde{h}_{1,2}$. We must impose that $\tilde{h}_1$ vanishes on the interval $[-l^2_{1},0]$ and $\tilde{h}_2$ on the interval $[k^2_1,\infty]$. Translating this condition to the $w$ plane means that $\tilde{h}_1$ vanishes on $[0,il_{1}]$
while $\tilde{h}_2$ vanishes on $[\infty,-k_1]$. We get
\begin{equation}\label{eq:general dual harm}
\begin{split}
\tilde{h}_1 = &  4 \mathrm{Re}(w) + 2i \sum^{m}_{b=1} \tilde{d}_b \ln \left[ \frac{(l_b - i w)(l_b - i \bar{w})}{(l_b + i w)(l_b + i \bar{w})}\right] ,\\
\tilde{h}_2 = & 4 \mathrm{Im}(w) - 2i \sum^{n}_{a=1} d_a \ln  \left[\frac{(k_a + w)(k_a - \bar{w})}{(k_a - w)(k_a + \bar{w})} \right] - 4 \pi \sum^{n}_{a=1} d_a .\\
\end{split}
\end{equation}
Consider $\tilde{h}_1$ first. Each term in the sum is proportional to $\mathrm{Im}[\ln(l_b-iw)-\ln(l_b+iw)]$ and thus vanishes on the interval $[0,il_b]$ where the logarithms have the same imaginary part \footnote{We take all the branch cuts to lie outside of the second quadrant, and choose the principal branch for all the logarithms.}. Hence the condition is satisfied. For $\tilde{h}_2$, by the same argument, the sum of logarithms vanishes on $[-k_n,0]$. Going along the negative real axis we jump over $n$ discontinuities, accumulating a contribution of $4 \pi d_a$ from each one of them, and thus we obtain the desired result on $[\infty,-k_1]$.

To compute the 5-form flux we need the value of $C_{(2)}$ ($=-2\tilde{h}_1 \hat{e}^{67}+\ldots$) at the position of the NS5-branes. Expanding $\tilde{h}_1$ around $w=-k_a$ we find
\begin{equation}
C_{(2)} = 8(k_a + 2 \sum^{m}_{b=1} \tilde{d}_b \arctan \left( \frac{k_a}{l_b}\right) ) + \cdots.
\end{equation}
Likewise, expanding $B_{(2)}$ ($=2\tilde{h}_2 \hat{e}^{45}+\ldots$) near the D5-branes at $w=il_b$ we find
\begin{equation}
B_{(2)} = 8(l_b - 2 \sum^{n}_{a=1} d_a \arctan \left( \frac{k_a}{l_b}\right)) + \cdots.
\end{equation}
The 5-form flux coming form the $a$'th stack of NS5-branes is given by
\begin{equation}
\int_{\Sigma_5^a} {\cal F}_1 = 8 \pi n_a (k_a + 2 \sum^{m}_{b=1} \tilde{d}_b \arctan \left( \frac{k_a}{l_b}\right)),
\end{equation}
and the flux coming from the $b$'th stack of D5-branes is
\begin{equation}
\int_{\Sigma_5^b} {\cal F}_2 = 8 \pi m_b (l_b - 2 \sum^{n}_{a=1} d_a \arctan \left( \frac{k_a}{l_b}\right)).
\end{equation}
The sum of all these fluxes exactly cancels the 5-form flux from $u=\infty$, as expected.

Note that the 5-form flux going into NS5-brane singularities is bounded from below by zero ($n_a, k_a, {\tilde d}_b, l_b > 0$), with the bound attained in the limit $k_a \to 0$. Similarly, the quantized 5-form flux per D5-brane going into D5-brane singularities is bounded from below by minus the total number of NS5-branes $\#_{NS5} = \sum_a n_a = 32\pi^2 \sum_a d_a$ (since $m_b>0$, $\arctan(x)< \pi/2$), with the bound attained in the limit $l_b \to 0$.
If we take the other
choice for the topology of $\Sigma_1$ and $\Sigma_2$, we would obtain a different value for these charges, but one that
would still be conserved; the difference between the two choices is a shift in the 5-form flux coming
from all D5-brane singularities by a constant times the D5-brane charge there times the total number of NS5-branes, and a shift in the opposite direction of the 5-form flux coming from all NS5-brane singularities, by a constant times the NS5-brane charge there times the total number of D5-branes.

The boundary conditions corresponding to configurations of D3-branes ending on both D5-branes and NS5-branes were also classified in \cite{Gaiotto:2008sa,Gaiotto:2008ak}. In this case one has to be careful about
the fact that the number of D3-branes ending on each 5-brane is not well-defined, since this changes when we move an NS5-brane past a D5-brane. However, it was shown in \cite{Gaiotto:2008ak} that if one ``regulates'' a brane configuration for
D3-branes ending on D5-branes and NS5-branes by slightly separating the 5-branes along the $z$ direction, then one can define a ``linking number'' \cite{Hanany:1996ie} associated with each 5-brane, which does not change when the
branes are moved around. The possible boundary conditions are then in one-to-one correspondence with the list of linking numbers associated with the D5-branes and the NS5-branes. The linking number $L_b$ associated with a D5-brane was defined in \cite{Gaiotto:2008ak} as the net number of D3-branes ending on it from the right (namely, the number of D3-branes ending on it from the right, minus the number of D3-branes
ending on it from the left), plus the number of NS5-branes on its left (=at smaller values of $z$). Similarly, the linking number $K_a$ associated with an NS5-brane was defined as the net number of D3-branes ending on it from the right, plus the number of D5-branes on its left. As discussed in \cite{Gaiotto:2008ak}, both linking numbers obey $L_b > 0$, $K_a > 0$.

One may hope that these linking numbers would correspond to the 5-form fluxes that we defined above for each 5-brane stack (divided by the number of 5-branes in that stack), since these should be related to the numbers of D3-branes ending on each 5-brane, but this is clearly not correct. For one thing, we had two different definitions of the 5-form, and it is not clear which one should map to the linking numbers; another issue is that the linking numbers defined in \cite{Gaiotto:2008ak} do not sum to the total number of D3-branes, but rather to that number plus the total number of D5-branes $\#_{D5}$ times the total number of NS5-branes $\#_{NS5}$. However, it is easy to see how to correct both problems. An equally natural definition of the linking number for D5-branes in some brane configuration is by taking ${\tilde L}_b$ to be the net number of D3-branes ending on it from the right, minus the number of NS5-branes on its right. This simply differs from the previous definition $L_b$ by subtracting from it $\#_{NS5}$. Similarly, one can define a different linking number ${\tilde K}_a$ for NS5-branes, to be the net number of D3-branes ending on it from the right, minus the number of D5-branes on its right. This differs from the previous
definition $K_a$ by subtracting from it $\#_{D5}$ \footnote{Of course we could also shift the linking
numbers by other multiples of $\#_{D5}$ and $\#_{NS5}$, but the two definitions described here are the
simplest and most natural ones, and they are quantized to integer values, unlike the original definition given in \cite{Hanany:1996ie}.}. Now, if we choose to characterize the D5-branes by the linking number ${\tilde L}_b$, and the NS5-branes by the linking number $K_a$, then these linking numbers (which still uniquely characterize a given boundary condition) sum to the total number of D3-branes, and we claim that they can be identified with the 5-form fluxes that we found above, for the first choice of the topology of $\Sigma_1$ and $\Sigma_2$. Namely, we identify
\be
K_a = 32\pi (k_a + 2 \sum^{m}_{b=1} \tilde{d}_b \arctan \left( \frac{k_a}{l_b}\right)),\qquad
{\tilde L}_b = 32 \pi (l_b - 2 \sum^{n}_{a=1} d_a \arctan \left( \frac{k_a}{l_b}\right)).
\e
As a check of this claim, note that the linking numbers defined in this way obey $K_a > 0$, ${\tilde L}_b > -\#_{NS5}$, which is precisely the same as the lower bounds we found above. We could also choose the linking numbers $L_b$ for the D5-branes, and ${\tilde K}_a$ for the NS5-branes. With this choice the linking numbers also sum to the total number of D3-branes, and they would precisely match with the 5-form fluxes that we would find using the second choice for the topology of $\Sigma_1$ and $\Sigma_2$ above. Thus, we find a precise matching between the classification of our supergravity solutions, and the supersymmetric boundary conditions for D3-branes ending on 5-branes classified in \cite{Gaiotto:2008ak}.

Our general solution for the D3-branes ending on 5-branes is written in terms of $2g$ physical parameters (up to $SL(2,\mathbb{R})$ transformations of type IIB supergravity). These are the $n$ parameters $\{n_a\}$ that count the number of NS5-branes in each stack, the $n$ parameters $\{k_a\}$ that are related to the number of D3-branes ending on each of them, the $m=g-n$ parameters $\{m_b\}$ that count the number of D5-branes in each stack, and the $m$ parameters $\{l_b\}$ that are related to the number of D3-branes ending on them. When the number of D3-branes ending on each 5-brane in the $c$'th stack is equal to the number of D3-branes ending on each 5-brane in the $(c+1)$'th stack, the two stacks come together $k_c=k_{c+1}$ ($l_c=l_{c+1}$) and the solution reduces to the genus $g-1$ case with $n-1$ ($m-1$) stacks of NS5-(D5)branes and $m$ ($n$) stacks of D5-(NS5)branes. The fact that the 5-branes are separated in the $u$-plane and ordered along the boundary according to the number of D3-branes ending on them is natural, since this number controls
 the bending of the 5-branes once back-reaction is taken into account; see, for example, figure 11 of \cite{Gaiotto:2008sa}.

Of course, the gravity solutions are only weakly curved when the number of D3-branes $N$ is large, and also when the number of 5-branes $m_b,n_a$ in each stack is large, $m_b,n_a \gg 1$; the solutions for small values of $m_b$ or $n_a$ include highly curved 5-brane ``throats''. If we take the large $N$ limit while keeping the ratios of the positions of the singularities in the $u$-plane
fixed, the number of 5-branes scales as $\sqrt{N}$ (and also the number of D3-branes ending on each
5-brane scales as $\sqrt{N}$). This is the natural scaling in gravity, since then both the radius of
the $S^5$ in the asymptotic $AdS_5\times S^5$ region, and the radius of the $S^3$ in the 5-brane throats in units of the string scale, scale as $N^{1/4}$. But we can also take a different large $N$ limit keeping the numbers of 5-branes
fixed, and as long as this number is large, our solutions are still weakly curved.

\section{One-point functions of chiral operators}\label{sec:one-point}

We next turn to the computation of field theory observables in the backgrounds
described in \S\ref{sec:endingbranes}. The simplest possible observables are one-point functions. In
a conformal field theory without a defect/boundary these have to vanish, but in
a conformal field theory on a half-line $z > 0$ with boundary conditions
preserving the lower-dimensional conformal symmetry, scalar primary operators
${\cal O}$ of dimension $\Delta$ are allowed to have one-point functions
$\langle {\cal O} \rangle = c / z^{\Delta}$ \cite{McAvity:1995zd}. (If we view our solutions as describing the
${\cal N}=4$ SYM theory on $AdS_4$, this corresponds to a constant vacuum expectation value
of ${\cal O}$ on $AdS_4$.)

In our case we have the 4d ${\cal N}=4$ SYM theory living on a half-line.
The boundary conditions break the $SU(4)$ global symmetry of this theory to
$SO(4) \simeq SU(2)\times SU(2)$, so only operators that are singlets of $SO(4)$
are allowed to have one-point functions. What are the lowest dimension
operators that are allowed to have one-point functions? The lowest-dimension
operator related by AdS/CFT to the metric, which corresponds in the bulk to
a combination of the trace of the metric on $AdS_5$, its trace on $S^5$, and
the 5-form field, is a scalar operator of dimension $\Delta=2$ in the ${\bf 20'}$
representation of $SU(4)$ \cite{Kim:1985ez}. This representation contains one singlet
of $SO(4)$. If we denote the three adjoint scalar fields
corresponding to the motion of the D3-branes along the D5-branes by $X_i$ ($i=1,2,3$),
and the three fields corresponding to the motion along the NS5-branes by $Y_i$ ($i=1,2,3$),
then it is given by ${\cal O}_2 = N {\rm tr}(X_1^2+X_2^2+X_3^2-Y_1^2-Y_2^2-Y_3^2)$.
The lowest-dimension scalar operator coming from the 2-form fields is a dimension
$\Delta=3$ complex scalar operator in the ${\bf 10}$ representation of $SU(4)$;
again this contains one singlet of $SO(4)$. Denoting the gauginos of the ${\cal N}=4$
SYM theory by $\lambda_a$ ($a=1,2,3,4$), the form of this operator is schematically
${\cal O}_3 = N {\rm tr}(\lambda_a \lambda_a + X_1 [X_2, X_3] + i Y_1 [Y_2, Y_3])$
(we assume that the kinetic terms of all ${\cal N}=4$ SYM fields are proportional to $1/g_{YM}^2$).
Finally, the lowest-dimension scalar operator coming from the dilaton-axion sector
is a dimension $\Delta=4$ complex singlet operator, whose real part takes the schematic form
${\cal O}_4 = N {\rm tr}(F_{\mu \nu}^2 + {\rm fermions} + \sum_{i < j} [X_i, X_j]^2 +
\sum_{i < j} [Y_i, Y_j]^2 + \sum_{i,j} [X_i, Y_j]^2)$.

Using the gravity solutions, we can compute the one-point functions of these operators
(and any other chiral operators) in the limit of large $N$ and large 't Hooft coupling.
To do this, we need to consider the
behavior of the background fields close to the boundary of our solutions at $u=\infty$,
where the solution is approximately $AdS_5\times S^5$.
In terms of the coordinate $v=-1/u$, the holomorphic differentials \eqref{eq:nholodiff} have the following asymptotic expansion near $v=0$ :
\be
\begin{split}
\partial h_1 = & - i \left( \gamma_1 \frac{1}{v^{3/2}} + \delta_1 \frac{1}{\sqrt{v}} + \eta_1 \sqrt{v} \right)dv + O(v^{3/2}),\\
\partial h_2 = & -  \left( \gamma_2 \frac{1}{v^{3/2}} + \delta_2 \frac{1}{\sqrt{v}} + \eta_2 \sqrt{v} \right) dv+ O(v^{3/2}),\\
\end{split}
\e
where the values of $\gamma_a, \delta_a, \eta_a$ depend on the specific solution.
 %labels the different solutions.

In terms of real coordinates
\be
v=e^{-2(x+iy)}, \quad - \infty \le x \le \infty, \quad  0 \le y \le \pi/2,
\e
the asymptotic region $v \rightarrow 0$ maps to $x \rightarrow \infty$. The metric factors up to next-to-leading order are
\be
\begin{split}
\rho^2 = & 2 \sqrt{2 |\Delta|} + \frac{\sqrt{2 |\Delta|}}{\gamma_1 \gamma_2} \left\{ (\gamma_1 \delta_2 + \gamma_2 \delta_1) \cos (2y) + 2\frac{\Omega}{\Delta}\cos(2y) \right\} e^{-2x} + O(e^{-4x}),\\
f_1^2 = & 8 \sqrt{2 |\Delta|} \cos^2 (y)  + 4 \frac{\sqrt{2 |\Delta|}}{\gamma_1 \gamma_2} \left\{(\gamma_1 \delta_2 + \gamma_2 \delta_1) [-2-\cos (2y)] + 2\frac{\Omega}{\Delta}\cos(2y) \right\}\cos^2 (y) e^{-2x} + O (e^{-4x}),\\
f_2^2 = & 8 \sqrt{2 |\Delta|} \sin^2 (y)  + 4 \frac{\sqrt{2 |\Delta|}}{\gamma_1 \gamma_2} \left\{ (\gamma_1 \delta_2 + \gamma_2 \delta_1) [2-\cos (2y)] + 2\frac{\Omega}{\Delta}\cos(2y) \right\} \sin^2 (y) e^{-2x} + O (e^{-4x}),\\
f_4^2 = & 8 \frac{|\gamma_1| |\gamma_2|}{\sqrt{2 |\Delta|}} e^{2x} + 4 \frac{|\gamma_1| |\gamma_2|}{\gamma_1 \gamma_2 \sqrt{2 |\Delta|}} \left\{ [2\Delta+(\gamma_1 \delta_2 +\gamma_2 \delta_1) \cos(2y)] - 2\frac{\Omega}{\Delta}\cos(2y) \right\} + O(e^{-2x}),\\
\end{split}
\e
where $\rho^2$ is the coefficient of $4(dx^2+dy^2)$ and we introduced the notation
\be
%\begin{split}
\Delta\equiv \gamma_1 \delta_2 - \gamma_2 \delta_1,\qquad\qquad
\Omega\equiv (\gamma_1)^2 \gamma_2 \eta_2-(\gamma_2)^2 \gamma_1 \eta_1.
%\end{split}
\e

So far we've been working in a ``conformal gauge", in which the residual diffeomorphism invariance of the supergravity solution consists of conformal transformations of the Riemann surface $\Sigma$. In order to easily read off the supergravity prediction for the vacuum expectation value (VEV) of the corresponding operators of the dual CFT, the solution has to be rewritten in the de Donder-Lorentz gauge, in which the contribution from the $SO(6)$ singlet spherical harmonic to the Kaluza-Klein expansion of the metric compactified on $S^5$ vanishes \cite{Kim:1985ez}. This is achieved by the diffeomorphism
\be \la{eq:diffeo}
e^{2x} \rightarrow \frac{1}{2}\frac{ |\Delta|}{|\gamma_1| |\gamma_2|} \left(e^{2x}  - \frac{1}{2} a_y \cos(2y)\right), \qquad \sin^2(y) \rightarrow \sin^2(y) \left(1+ a_y \cos^2(y) e^{-2x} \right),
\e
with
\be
a_y = - 4 \frac{|\gamma_1| |\gamma_2|}{\gamma_1 \gamma_2} \frac{1}{|\Delta|}(\gamma_1 \delta_2 +\gamma_2 \delta_1).
\e

The metric then becomes
\be\label{asymmetric}
%\begin{split}
ds^2= 8 \sqrt{2 |\Delta|} (ds^2_{AdS_5}+ds_{S^5}^2)+ 8 \sqrt{2 |\Delta|} \delta \zeta \cos(2y)  e^{-2x} \left(ds_{S^5}^2 + dx^2 - \frac{1}{4} e^{2x}ds^2_{AdS_4}\right) + O(e^{-4x}),
% \end{split}
 \e
 where
 \be
 \delta \zeta = \frac{1}{ |\Delta|}  \frac{|\gamma_1| |\gamma_2|}{\gamma_1 \gamma_2} \left[ -3 (\gamma_1 \delta_2 + \gamma_2 \delta_1)+ 2\frac{\Omega}{\Delta}\right].
 \e
 The dilaton and the functions defining the 2-form potentials up to next-to-leading order are
 \be
 \begin{split}
e^{\Phi}= & \left|\frac{\gamma_2}{\gamma_1}\right|+\frac{1}{2}\left|\frac{\gamma_2}{\gamma_1}\right| \frac{\Delta}{ (\gamma_2 \gamma_1)^2 } \left[3  (\gamma_1 \delta_2 + \gamma_2 \delta_1) - 2\frac{\Omega}{\Delta} \right] e^{-4x} + O(e^{-6x}),\\
b_1= & \frac{32}{3} \frac{1}{\sqrt{2|\Delta|}} \frac{\Delta}{|\Delta|}\left|\frac{\gamma_2}{\gamma_1}\right|^{\frac{1}{2}} \left[ 3 (\gamma_1 \delta_2 + \gamma_2 \delta_1) - 2\frac{\Omega}{\Delta}\right] \cos^3(y) e^{-3x} + O(e^{-5x}), \\
b_2= & \frac{32}{3} \frac{1}{\sqrt{2|\Delta|}}  \frac{\Delta}{|\Delta|} \left|\frac{\gamma_1}{\gamma_2}\right|^{\frac{1}{2}} \left[ 3 (\gamma_1 \delta_2 + \gamma_2 \delta_1)- 2\frac{\Omega}{\Delta}\right] \sin^3(y) e^{-3x} + O(e^{-5x}). \\
\end{split}
\e

For the special case of D3-branes ending on $n=g$ stacks of NS5-branes, the constants describing the asymptotic behavior of the real harmonic functions $h_1$ and $h_2$ are
\be
\begin{split}
\gamma_1&= i, \quad \quad \gamma_2  = i,\\
\delta_1&=  0, \quad \quad \delta_2  = i \sum_{a=1}^g (\beta_a-k_a^2),\\
\eta_1&= 0, \quad \quad \eta_2 =  -i \sum_{c \ne a}^g [\frac{1}{2}(\beta_c \beta_a+k_c^2k_a^2)-\beta_c k_a^2]-i \sum_{a=1}^g (k_a^4-\beta_a k_a^2), \\
% = \sum_{a \ne b}^g \frac{1}{2}(\beta_a-k_a^2)(\beta_b-k_b^2) - \sum_{a=1}^g k_a^2(\beta_a-k_a^2);\\
\end{split}
\e
such that $\Delta = i \delta_2$ and $\Omega =-i \eta_2$. The number of D3-branes ending on the 5-branes is thus $N = 8 (4\pi)^3 \sum_{a=1}^g (\beta_a - k_a^2)$. We can then write
\be
\begin{split}\label{asymvalues}
\delta \zeta = & \frac{1}{ |\delta_2|^2 }[3 (\delta_2)^2- 2 i \eta_2],\\
e^{\Phi}= & 1 - \frac{1}{2} [3 (\delta_2)^2- 2 i \eta_2] e^{-4x} + O(e^{-6x}),\\
b_1= &  \frac{16 \sqrt{2}}{3} \frac{1}{|\delta_2|^{\frac{3}{2}}}[3 (\delta_2)^2- 2i \eta_2] \cos^3(y) e^{-3x} + O (e^{-5x}),\\
b_2= & \frac{16 \sqrt{2}}{3} \frac{1}{|\delta_2|^{\frac{3}{2}}} [3 (\delta_2)^2- 2 i \eta_2] \sin^3(y) e^{-3x} + O (e^{-5x}).
\end{split}
\e
Expressed (implicitly) in terms of the numbers of 5-branes (through $\{\beta_a\}$), and the numbers of D3-branes ending on
each 5-brane, we can read off from \eqref{asymmetric}, \eqref{asymvalues} the following expectation values for ${\cal O}_{2,3,4}$ (up to an overall normalization of each operator that we do not carefully fix here)\footnote{Note that generally the one-point functions of operators are not simply related to the coefficients of the normalizable modes of the corresponding fields near the boundary of $AdS_5$, but have additional contributions involving the normalizable modes of other fields; see, for instance, \cite{Skenderis:2006uy}. However, for the specific operators that we discuss here, the additional contributions are absent.} :
\be
\begin{split}\label{gravityvevs}
\langle {\cal O}_2 \rangle \propto &
%\frac{\sqrt{2}}{\pi}N^{\frac{1}{2}}\left[4-\frac{\sum_{a=1}^g (\beta_a^2 - k_a^4)}{\sum_{a,b=1}^g (\beta_a - k_a^2)(\beta_b - k_b^2)}\right]
\left[ \frac{N^2}{16(4\pi)^6} - \sum_{a=1}^g (\beta_a^2 - k_a^4) \right]
\frac{1}{z^2},\\
\langle {\cal O}_3 \rangle \propto &
%\frac{2\sqrt{2}}{3 \pi}N^{\frac{1}{2}}\left[4-\frac{\sum_{a=1}^g (\beta_a^2 - k_a^4)}{\sum_{a,b=1}^g (\beta_a - k_a^2)(\beta_b - k_b^2)}\right]
\left[ \frac{N^2}{16(4\pi)^6} - \sum_{a=1}^g (\beta_a^2 - k_a^4) \right]
\frac{1}{z^3},\\
\langle {\cal O}_4 \rangle \propto &
%\frac{1}{2^{13}\pi^4} N^2 \left[4-\frac{\sum_{a=1}^g (\beta_a^2 - k_a^4)}{\sum_{a,b=1}^g (\beta_a - k_a^2)(\beta_b - k_b^2)}\right]
\left[ \frac{N^2}{16(4\pi)^6} - \sum_{a=1}^g (\beta_a^2 - k_a^4) \right]
\frac{1}{z^4}.\\
\end{split}
\e
For the special case of D3-branes ending on $m=g$ stacks of D5-branes, the expectation values are the same up to the replacement $\beta_a \rightarrow -\alpha_b$ and $k_a \rightarrow l_b$.

Note that the simplest large $N$ limit involves scaling all special points on the real axis as $N$,
namely $\beta_a \propto N$ and $k_a \propto \sqrt{N}$. In this limit the number of 5-branes scales
as $\sqrt{N}$, and the expectation values above scale as $N^2$ (which is the standard normalization
of all correlation functions in the large $N$ limit). If we want the number of 5-branes to be of
order $N$, we need to leave $k_a$ fixed in the large $N$ limit, but the one-point functions still scale
as $N^2$. On the other hand, if we want the number of 5-branes to remain of order one, we need to
take $k_a \propto N$, $(\beta_a - k_a^2) \propto N$ in the large $N$ limit; in this limit the one-point
functions \eqref{gravityvevs} scale as $N^3$.

We can compare \eqref{gravityvevs} to the same expectation values at weak coupling. As reviewed
above, the weak coupling boundary conditions were discussed in \cite{Gaiotto:2008sa,Gaiotto:2008ak}. For D3-branes ending
on NS5-branes these boundary conditions involve a strongly coupled 3d SCFT living at
$z=0$, so we do not know how to compute anything. However, for D3-branes ending purely on
D5-branes, the boundary conditions are given by $X_i = \tau_i / z$, where $\tau_i$
is some $N$-dimensional representation of $SU(2)$ ($[\tau_i, \tau_j] = i \epsilon_{ijk} \tau_k$),
and we can use this to compute the
expectation values of our operators in the weak coupling limit. Our solutions involve $g$
stacks of D5-branes, with $m_b$ D5-branes in each stack, and $L_b$ D3-branes ending on each
5-brane in the $b$'th stack, and we identified them above with the $N$-dimensional
representation of $SU(2)$ that has $m_b$ blocks of size $L_b\times L_b$.

To compute $\langle {\cal O}_2 \rangle$ at weak coupling, we thus need to compute
${\rm tr}(X_1^2+X_2^2+X_3^2)$ in this representation. This is proportional to $\sum_b m_b C_{L_b}$, where $C_{L_b}$
is the second Casimir of the $L_b$-dimensional representation of $SU(2)$, equal to
$C_{L_b} = (L_b^2 - 1) / 2$. Thus, in the large $L_b$ limit in which our solutions are valid we expect
$\langle {\cal O}_2 \rangle = N (\sum_b m_b L_b^2) / z^2$, up to a multiplicative constant that
is independent of $m_b, L_b$. In fact, given the expressions
above for ${\cal O}_{3,4}$, it is easy to see using the $SU(2)$ algebra that they are
also proportional to precisely the same expression, just with a different power of $z$. One can check that
these results do not agree with the strong coupling results \eqref{gravityvevs} computed above, indicating that
the one-point functions of these operators have a non-trivial dependence on the 't Hooft
coupling. In fact, when the number of 5-branes is of order $\sqrt{N}$, we even find a different
power of $N$ at weak and strong coupling; in this case at weak coupling the one-point functions scale as
$N^{5/2}$. On the other hand, when the number of 5-branes is of order one we find weak-coupling
one-point functions of order $N^3$, and when it is of order $N$ we find weak-coupling one-point
functions of order $N^2$, which is similar to the strong coupling behavior (but the precise
dependence on the number of D3-branes ending on each 5-brane stack is different).

It is curious that both at weak coupling and at strong coupling, all three operators
have the same expectation value (up to an overall constant that we did not fix, but the dependence of all three operators on the numbers of D3-branes ending on each 5-brane is the same); perhaps
this indicates some non-renormalization theorem for ratios of expectation values. It would be interesting to try to compute these one-point functions exactly as
a function of the 't Hooft coupling; perhaps this can be done, like similar computations,
using integrability or localization methods.

\section{Summary and conclusions}\label{summary}

In this paper we used the results of \cite{D'Hoker:2007xy,D'Hoker:2007xz} to construct gravitational
 duals for the ${\cal N}=4$ SYM theory on ${\mathbb R}^{2,1}$ times a half-line (or on $AdS_4$) with various boundary conditions that preserve half of the supersymmetry, describing the near-horizon limit of D3-branes ending on 5-branes. We obtain an explicit closed form for these solutions, given by plugging \eqref{eq:general harm} into the equations of section \ref{dhetalreview}, and we find a one-to-one mapping between our solutions and the boundary
 conditions for D3-branes ending on 5-branes, classified in \cite{Gaiotto:2008sa,Gaiotto:2008ak}. Assuming that the classification of solutions in \cite{D'Hoker:2007xy,D'Hoker:2007xz} is complete, we present the most general solutions of this type.
These should correspond to the most general supersymmetric boundary conditions of ${\cal N}=4$ SYM that have a supergravity (with 5-branes) approximation for some range of their parameters; there can also be other types of boundary conditions that never have a purely supergravity description, such as the orientifold/orbifold boundary conditions discussed in \cite{Aharony:2010ay}.

A simple generalization of the solutions we find (which goes beyond supergravity)
involves adding $M$ D3-branes sitting at the point $u=0$ where the two two-spheres
go to zero size. This gives a generalized boundary condition with an extra $U(M)$
global symmetry, coming from the gauge symmetry on the D3-branes; in the field theory
this comes from $M$ additional charged matter fields living on the
boundary. We can think of the new boundary condition in the language
of the brane construction as starting from a solution with $M$ semi-infinite D3-branes
on the other side of the 5-branes, but taking a limit where the $3+1$ dimensional
gauge theory on these D3-branes decouples, leaving behind a global symmetry. Such a
decoupling limit involves taking the gauge coupling on these D3-branes to zero. In the
brane construction we cannot really do this since the string coupling on both stacks
of semi-infinite D3-branes is the same, but in the solutions of \cite{D'Hoker:2007xy,D'Hoker:2007xz} there are independent string coupling parameters
for the two stacks of semi-infinite D3-branes (as in the ``Janus solutions'') so such a
limit is possible. Naively we would describe such a limit by starting with an extra
$AdS_5$ singularity at $u=0$ and taking the asymptotic string coupling down the $AdS_5$
throat to zero, but we claim that the limiting solution is simply described by putting
$M$ D3-branes at $u=0$. Note that the string scale in our solutions is finite, so we
cannot replace the D3-branes by an $AdS_5\times S^5$ ``throat''. The precise description of the new boundary conditions in gauge
theory can be derived along the lines of \cite{Gaiotto:2008sa,Gaiotto:2008ak}, just
adding $M$ extra semi-infinite D3-branes (with vanishing gauge coupling on their
worldvolume). This gives $M$ extra charged fields under the last gauge group in the
quiver diagram. One can also obtain such fields by adding $M$ extra 5-branes, so the
solutions described in this paragraph are not independent of the general solutions we
described above, but should be thought of as a different way to describe a limit of the general solutions in which the linking number of some 5-branes is very small. This alternative description could be more
useful for some range of parameters.

There are many remaining open questions.
In this paper we only studied the solutions of \cite{D'Hoker:2007xy,D'Hoker:2007xz} that have no $u_a$
points (zeros of $\partial h_{1,2}$) in the middle of the Riemann manifold $\Sigma$,
since solutions with such points appear to have conical singularities. It
would be interesting to study further the solutions with $u_a$ points,
to see if in string theory there is some way to resolve their singularities.

All of our solutions involve regions which look like NS5-branes and/or D5-branes wrapped on $AdS_4\times S^2$. In these regions the dilaton blows up (for NS5-branes) and supergravity breaks down, which is not surprising since there are many light fields hiding there that are not seen in supergravity (in particular, for $m$ 5-branes there are $U(m)$ gauge fields living on $AdS_4\times S^2$). The solutions near $m$ NS5-branes involve a ``throat'' region where the radius of curvature (in the string frame) is $\sqrt{m}$ times the string scale, so for small $m$ stringy corrections to supergravity are important. Note that from the point of view of our solutions the ``natural'' scaling at large $N$ (where $N$ is the number of D3-branes) is to have the number of 5-branes in each stack scale as $\sqrt{N}$, since only in this case the supergravity solution scales uniformly when we take large $N$. However, our solutions are also well-behaved (away from the 5-branes) when $m$ is large and fixed in the large $N$ limit, and only in the fixed $m$ case do we expect to have a standard 't Hooft large $N$ limit (in which the number of gauge-invariant operators remains fixed at large $N$). For NS5-branes in flat space there is a well-known string theory description of the corresponding ``throat'' using an exact worldsheet CFT, and it would be interesting to see if this can be extended to the case of 5-branes on $AdS_4\times S^2$. For 5-branes in flat space one can resolve the strong coupling region by slightly separating the 5-branes in specific ways (as in, for instance, \cite{Giveon:1999px}), and it would be interesting to see if this can be done also in our case, by splitting the 5-branes along the real axis in the $u$-plane. A particularly interesting case is that of a single NS5-brane; the general boundary conditions involving NS5-branes include non-trivial 3d SCFTs on the boundary, but the single NS5-brane corresponds just to simple Neumann/Dirichlet boundary conditions for all the fields of the ${\cal N}=4$ SYM theory, so it is the only case with NS5-branes that has a weakly coupled description. From the gravity point of view, we get also in this case a highly curved ``throat'', but since in this case there is no non-Abelian gauge symmetry hidden in the ``throat'', it is plausible that this ``throat'' has a smooth resolution in string theory with no strong coupling region. This issue deserves further study.

There are many computations that can be done using the solutions we find; in this paper we only computed a few one-point functions of chiral operators of the ${\cal N}=4$ SYM theory, and found that they do not agree with the weak coupling results. It would be interesting to analyze the behavior of these one-point functions as a function of the 't Hooft coupling, to see if it can be found exactly. It would also be interesting to compute other observables, and to see if there are any observables in these theories that are protected by supersymmetry (independent of the 't Hooft coupling).
%; such observables could be useful for completing the precise mapping of our solutions to field theory. 
While the one-point functions in such backgrounds are uniquely determined (up to a constant) by the conformal symmetry, two-point functions are not \cite{McAvity:1995zd}, and it would be interesting to compute them and to see what they teach us about these theories. It is particularly interesting to compute the spectrum of our solutions, which maps to the spectrum of anomalous dimensions of 3d ``boundary operators'' in the field theory; this computation was recently discussed in \cite{Bachas:2011xa} for a more general class of solutions, but it is beyond the scope of this paper. One could also analyze the spectrum of states which are not part of supergravity, such as (D)-strings stretched between 5-brane stacks, or branes wrapping non-trivial cycles in our solutions. There are also states coming from the fields living on the wrapped 5-branes; the states coming from the massless fields on the 5-branes wrapping $AdS_4\times S^2$, which are in short representations of $OSp(4|4)$, were classified in \cite{DeWolfe:2001pq}.

There are many possible generalizations of our solutions, but most of the interesting ones involve configurations with less supersymmetry, so they would be harder to construct. This includes in particular the brane configurations of D3-branes stretched between 5-branes, and of D4-branes ending on (or stretched between) 5-branes, whose construction was one of the main motivations for this work. There is one case which involves the same amount of supersymmetry, which is that of M2-branes ending on M5-branes, and it would be interesting to generalize the analysis of our paper to this case using the solutions of \cite{D'Hoker:2008wc,D'Hoker:2009my}. The field theory corresponding to this case was recently discussed in \cite{Berman:2009kj,Chu:2009ms,Berman:2009xd}.

For solutions that have both NS5-brane and D5-brane singularities adjacent to the
$AdS_5\times S^5$ singularity, one can also consider
a limit of our solutions in which the D3-brane flux in the single asymptotic
$AdS_5\times S^5$ region goes to zero. In this limit the $\alpha$ and $\beta$ points
adjacent to the $AdS_5\times S^5$ singularity approach this singularity.
This gives a solution which is a warped
product $AdS_4\times M_6$ with a manifold $M_6$ which is compact (except for 5-brane ``throats''); such a solution is dual to some
3d ${\cal N}=4$ superconformal theory, without any coupling to a 4d theory. Starting from a solution that has an interpretation as D3-branes
ending on D5-branes and NS5-branes, we can interpret this theory as
the low-energy theory on the D3-branes stretched between these
D5-branes and NS5-branes.

Finally, it would be interesting to generalize the solutions we find to finite temperature. Here there is a difference between considering our solutions as describing the ${\cal N}=4$ SYM theory on a half-line or on $AdS_4$, and the finite temperature generalization can be considered in both cases. In the first case it is clear that the asymptotic $AdS_5\times S^5$ region should be replaced by the near-extremal D3-brane solution, and it would be interesting to see how this is completed to the full geometry. The second case has richer dynamics, since (if we use global coordinates for $AdS_4$) it has a dimensionless parameter (the temperature in units set by the $AdS_4$ radius), and one expects (as discussed in \cite{Aharony:2010ay}) phase transitions as a function of this parameter. In this case there is always a trivial solution where we just periodically identify the (Euclidean) time direction of the $AdS_4$ factor in our solutions, and this trivial solution should be the dominant one at low temperatures, but at some point we expect a phase transition to a new solution with a horizon. It would be interesting to find and analyze these new solutions for the various boundary conditions we discuss in this paper.

%~~~~~~~~~~~~~~~~~~~~~~~~~~~~~~~~~~~~~~~~~~~~~~~
\subsection*{Acknowledgments}
\label{s:acks}
%~~~~~~~~~~~~~~~~~~~~~~~~~~~~~~~~~~~~~~~~~~~~~~

It is a pleasure to thank Costas Bachas, Francesco Benini, Cyril Closset, Stefano Cremonesi, John Estes, Daniel Jafferis, David Kutasov, Mukund Rangamani, Cobi Sonnenschein, Shimon Yankielowicz, and especially Don Marolf for useful discussions. We thank Nizan Klinghoffer for assistance with the figures.
This work was supported in part by the Israel--U.S.~Binational Science Foundation, by a research center supported by the Israel Science Foundation (grant number 1468/06), by the German-Israeli Foundation (GIF) for Scientific Research and Development, and by the Minerva foundation with funding from the Federal German Ministry for Education and Research.

%\appendix

\end{document}